\documentclass[iop]{emulateapj}

\usepackage{graphicx}
\usepackage{aas_macros}
\usepackage{amsmath}
\usepackage{amssymb}
\usepackage{graphics}
\usepackage{natbib}

\shorttitle{PRIMUS: The relationship between Star Formation and AGN accretion}
\shortauthors{Azadi et al.}
 \slugcomment{Accepted to \apj}

\begin{document}

\title{PRIMUS: The relationship between Star formation and AGN accretion}
\author{
Mojegan Azadi\altaffilmark{1}, 
James Aird\altaffilmark{2,3}, 
Alison L. Coil\altaffilmark{1}, 
John Moustakas\altaffilmark{4},
Alexander J. Mendez\altaffilmark{1},
Michael R. Blanton\altaffilmark{5}, 
Richard J. Cool\altaffilmark{6}, 
Daniel J. Eisenstein\altaffilmark{7}, 
Kenneth C. Wong\altaffilmark{8}, 
Guangtun Zhu\altaffilmark{9}
}
\altaffiltext{1}{Center for Astrophysics and Space Sciences, Department of Physics, University of California, 9500 Gilman Dr., La Jolla, San Diego, CA 92093, USA}
\altaffiltext{2}{Department of Physics, Durham University, Durham DH1 3LE, UK}
\altaffiltext{3}{Institute of Astronomy, University of Cambridge, Madingley Road,
Cambridge CB3 0HA, UK}
\altaffiltext{4}{Department of Physics and Astronomy, Siena College, 515 Loudon Road, Loudonville, NY 12211, USA}
\altaffiltext{5}{Center for Cosmology and Particle Physics, Department of Physics, New York University, 4 Washington Place, New York, NY 10003, USA}
\altaffiltext{6}{MMT Observatory, 1540 E Second Street, University of Arizona, Tucson, AZ 85721, USA}
\altaffiltext{7}{Harvard College Observatory, 60 Garden St., Cambridge, MA 02138, USA}
\altaffiltext{8}{Institute of Astronomy and Astrophysics, Academia Sinica, No.1, Sec. 4, Roosevelt Rd., Taipei 10617, Taiwan}
\altaffiltext{9}{Department of Physics \& Astronomy, Johns Hopkins University, 3400 N. Charles Street, Baltimore, MD 21218, USA}

\begin{abstract}

We study the evidence for a connection between active galactic nuclei (AGN) fueling and star formation by investigating the relationship between the X-ray luminosities of AGN and the star formation rates (SFRs) of their host galaxies. We identify a sample of 309 AGN with $10^{41}<L_\mathrm{X}<10^{44} $ erg s$^{-1}$ at $0.2 < z < 1.2$ in the PRIMUS redshift survey. We find AGN  in galaxies with a wide range of SFR at a given $L_X$.
We do not find a significant correlation between SFR and the observed instantaneous $L_X$ for star forming AGN host galaxies. However, there is a weak but significant correlation between the mean $L_\mathrm{X}$ and SFR of detected AGN in star forming galaxies, which likely reflects that $L_\mathrm{X}$ varies on shorter timescales than SFR. We find no correlation between stellar mass and $L_\mathrm{X}$  within the AGN population. Within both populations of star forming and quiescent galaxies, we find a similar power-law distribution in the probability of hosting an AGN as a function of specific accretion rate. Furthermore, at a given stellar mass, we find a star forming galaxy $\sim2-3$ more likely than a quiescent galaxy to host an AGN of a given specific accretion rate. The probability of a galaxy hosting an AGN is constant across the main sequence of star formation.
These results indicate that there is an underlying connection between star formation and the presence of AGN, but AGN are often hosted by quiescent galaxies. 

\end{abstract}


\section{Introduction}
\label{sec:intro}

It has been several decades since the first observations of galaxies with strong emission lines in their central regions \citep{seyfert1943nuclear} and the classification of such sources as Active Galactic Nuclei (AGN).
Since then, the range of observational phenomena associated with AGN has expanded to include 
sources classified based on a variety of X-ray, optical, infrared and radio criteria \citep[e.g.][]{antonucci1993unified}
and there have been numerous investigations into the physical nature of these AGN \citep[for a recent review see][]{alexander2012drives}.
It is now widely accepted that AGN activity is due to the presence of a supermassive black hole (SMBH) accreting gas and dust in the circumnuclear region, forming an accretion disk that ultimately powers the AGN activity.
Studies from recent decades have established that SMBHs reside in almost all galaxies with a bulge or spheroid component  \citep[e.g.][]{ kormendy1995inward, kormendy2011supermassive}, but what triggers AGN activity in some galaxies and not others is still a matter of debate. 
During the accretion process a tremendous amount of energy is released, 
a fraction of which can be injected into the circumnuclear region, the host galaxy or even the wider galactic environment in the form of electromagnetic or mechanical output. 
However, it is still unclear to what extent the injection of energy takes place and whether it has a strong effect on evolution of the host galaxy. 

Various observational investigations support the idea that there is a close connection between the growth of SMBHs and the growth of their host galaxies. These studies find correlations between the SMBH mass and the bulge stellar mass  \citep{magorrian1998demography} or velocity dispersion \citep[e.g.][]{ferrarese2000fundamental,gebhardt2000relationship,kormendy2011supermassive} of the host galaxy. This indicates that SMBHs and galaxies must, on average, grow together, but whether this indicates a causal connection between these processes remains unclear \citep[e.g.][]{peng2007mergers,jahnke2011non}. Furthermore, the global star formation rate (SFR) density of galaxies and the SMBH accretion rate density evolve similarly with redshift \citep[e.g.][]{boyle1998cosmological,silverman2008luminosity,aird2010evolution,assef2011mid}, indicating that the growth of galaxies and their central SMBHs are related in a global sense.
However, it is unclear whether the level of AGN activity itself and host 
galaxy properties such as SFR  are linked within individual galaxies.

A number of theoretical models and simulations suggest that AGN activity and star formation in galaxies are linked through a common cold gas inflow from galaxy mergers \citep[e.g.][] {di2005energy,hopkins2006relation, somerville2008semi}. Some observational studies find that galaxies with the highest SFRs are associated with merger events \citep[e.g.][]{shi2009role, kartaltepe2010multiwavelength, kartaltepe2012goods}. Due to the high fraction of quasars in merging systems, several authors have proposed that nuclear activity is also tightly connected to merger events \citep[e.g.][]{Sanders96,canalizo2001quasi,ivison2010}. However, recent studies find little to no connection between AGN activity and the incidence of merger events in galaxies with moderate luminosity AGN \citep[e.g.][]{schawinski2011hst,kocevski2012candels}, suggesting that secular processes such as turbulence and disk instabilities are more effective in enhancing nuclear accretion activity for AGN with moderate luminosities \citep[e.g.][]{mullaney2012goods, rosario2012mean}.

Studies of the locations of AGN host galaxies in the optical color-magnitude diagram are useful to investigate the role of AGN in the evolution of their host galaxies. Galaxies can be divided into two general populations in this diagram:  the blue cloud, consisting of predominantly star forming galaxies, and the red sequence, comprised mainly of quiescent, passively evolving galaxies \citep[e.g.][] {blanton2003galaxy, bell2003optical, baldry2004quantifying}.  
There is also a third population, known as the green valley, that includes galaxies in transition between the other two populations \citep[e.g.][]{mendez2011aegis}. Many studies have demonstrated that X-ray detected AGN from $0 < z < 1$ are not preferentially found in galaxies with the highest levels of star formation. Instead, they appear to be in the reddest part of the blue cloud, in the green valley, or on the red sequence  \citep[e.g.][]{nandra2007aegis,coil2009aegis,hickox2009host, georgakakis2011serendipitous}. 
There is also evidence that  high-luminosity AGN preferentially reside in luminous, massive, bulge-dominated galaxies \citep[e.g.][]{Schade2000,pagani2003host,dunlop2003quasars,floyd2004host}.
Taken at face value, these results seem to indicate that AGN may, via feedback to their host galaxy, extinguish star formation and could be responsible for the transition of their hosts from the blue cloud to the red sequence. 

However, \cite{silverman2009environments} and  \cite{xue2010color} showed that stellar mass selection biases have strong effects on these results and that the preference for green and red host galaxies is due to AGN generally being found in massive galaxies. \cite{xue2010color} showed that while the fraction of AGN increases in a population with more massive galaxies, in a sample with matched stellar mass host galaxies, AGN are equally likely to be found in any host population. \cite{aird2012primus} further showed that the observed prevalence of massive host galaxies is due to a selection effect related to the underlying distribution of specific accretion rates (AGN luminosity scaled relative to host stellar mass). Essentially, a more massive galaxy tends to host a more massive BH, which is easier to detect for a given specific accretion rate. Therefore, while AGN will preferentially be detected in massive galaxies, they actually reside in galaxies with a wide range of stellar mass.

Whether there is a preference for AGN to be found in star forming galaxies, once stellar mass-dependent biases are accounted for, remains an open question. \citet{aird2012primus} measured the distribution of accretion rates within AGN hosts in the blue cloud, green valley, and red sequence and found a similar accretion rate distribution in all populations.  They further found that there was a mild (factor $\sim2$) enhancement in the probability of a galaxy of a given stellar mass hosting an AGN for galaxies with blue or green rest-frame optical colors. \citet{bongiorno2012accreting} used specific SFR (estimated from fits to the optical--to--near-infrared spectral energy distributions) to split their galaxy sample into star forming and quiescent galaxies, finding no significant differences in the probability of hosting an AGN for galaxies in either population. More recently, \citet{hernan2014higher} found that AGN hosts have similar distributions of rest frame optical colors to inactive galaxies of the same stellar mass but are more likely to be hosted by galaxies with younger stellar populations. \citet{georgakakis2014investigating} also split AGN hosts into star forming and quiescent populations based on their  $U-V$ versus $V-J$ colors (which should be more robust to dust extinction than rest-frame optical colors) and found that the space density of star forming hosts is higher than for quiescent hosts, with some weak evidence for differences in the shape of the accretion rate distributions.

A number of recent studies have used \textit{Herschel} far infrared data, which provides a more robust tracer of the total SFR and is not impacted by dust extinction, to compare AGN hosts to the wider galaxy population. In \textit{Herschel}-detected populations, \citet{mullaney2012goods}, \cite{santini2012enhanced} and \citet{rosario2013nuclear} all found evidence for enhanced SFR in AGN hosts, compared to non-active galaxies of the same stellar mass, and argued that the bulk of moderate-luminosity AGN are hosted by normal star forming galaxies. However, \textit{Herschel} is only able to detect galaxies that are bright at far-infrared wavelengths and thus generally have high SFRs, making it difficult to measure the fraction of quiescent hosts and compare the SFRs of all AGN hosts to the full population of star forming galaxies.
However, given that the presence of dust can redden UV-optical colors, results that rely solely on optical colors may be biased.  
\cite{cardamone2010dust} found that dust reddening affects the colors of some star forming AGN host galaxies, pushing them to the green valley. 

In addition to determining whether AGN are more likely to reside in star forming or quiescent host galaxies, several authors have studied whether there is an overall correlation between the level of star formation and the level of nuclear activity, as traced by the X-ray luminosity, in individual galaxies. \cite{rovilos2012goods} found no evidence for a correlation in AGN with  $L_\mathrm{X}< 10^{43.5}$ erg s$^{-1}$ at $z<1$ but a significant correlation at higher X-ray luminosity at $z>1$, using a sample of X-ray detected AGN  in the \textit{Chandra} Deep Field--South (CDFS).
\cite{mullaney2012goods} use \textit{Herschel}-detected moderate luminosity
($L_\mathrm{X}=10^{42-44}$ erg s$^{-1}$) X-ray AGN in the CDFS and 
\textit{Chandra} Deep Field--North (CDFN) fields at $0<z<3$ and found no 
evidence of a correlation between SFR and X-ray luminosity, once the overall 
evolution of the average SFR with redshift is accounted for.
\cite{rosario2012mean} find similar results, using even larger AGN samples, 
including the COSMOS field.  However, \cite{rosario2012mean} did find a correlation in the most luminous AGN with $L_{AGN}\gtrsim10^{45}$ erg s$^{-1}$ at $z<1$, which they 
interpret as due to major merger events. In addition, \cite{mullaney2012hidden} found that the ratio of SMBH growth to SFR does not change with the stellar mass and redshifts at $z<2.5 $. Therefore, they suggest that rather than violent mergers, secular processes are responsible for both star formation and SMBH growth in majority of the galaxies with moderate nuclear activity.

There is some evidence that optically luminous AGN are found in galaxies with enhanced SFRs \citep[e.g.][]{floyd2012star}. 
However, \cite{page2012suppression} 
found that star formation was suppressed in their sample of luminous X-ray detected AGN at $1< z <3$ with spectroscopic redshifts and \textit{Herschel}/SPIRE 250 $\mu m$ detections in the CDFN. 
More recently, \cite{harrison2012no} found no sign of star formation suppression in powerful AGN hosts using a larger sample. Moreover, \cite{ harrison2012no} found that the average SFR in galaxies hosting AGN with luminosities in range of $10^{43}<L_\mathrm{X}<10^{45} $ erg s$^{-1}$ at $1< z <3$ is constant as a function of $L_\mathrm{X}$, consistent with results from \cite{mullaney2012goods} and \cite{rosario2012mean}. 

Studies of lower luminosity AGN have been carried out in the local Universe.
Using a sample of nearby Seyfert galaxies \cite{diamond2012relationship} found a strong correlation between the AGN luminosity and the SFR in the circumnuclear regions ($r <1$ kpc); however they found no correlation with the galaxy-wide SFR.  \cite{kauffmann2003host} studied optically-selected AGN in SDSS and found that 
strong nuclear activity is associated with younger stellar populations, indicative of higher levels of recent star formation. 

All of the above studies use instantaneous accretion rate to investigate correlations with SFR. However, a number of recent studies 
have instead focused on the \emph{average} accretion rate, averaging over both active and non-active galaxies, and SFR.  \cite{chen2013correlation} studied star forming galaxies detected by \textit{Herschel} in the Bo\"{o}tes field at $z<1$, including 34 X-ray and 87 MIR-detected AGN, and found that the average accretion rate is correlated with SFR.  They also compared their result with a small sample of 20 X-ray detected AGN in FIR bright galaxies from \cite{symeonidis2011herschel} at $z \sim$1 and found a consistent trend in both samples. However, when they corrected for the effects of flux limits on their results, along with the evolution of the X-ray luminosity function and average SFR with redshift, the correlation becomes weaker.  \cite{hickox2014black} presented a model where all star forming galaxies host an AGN and the average AGN luminosity is correlated with the SFR. This model explains the correlation between the average AGN luminosity and SFR seen by \citet{chen2013correlation}. 
They also use their model to investigate the correlation between the instantaneous accretion rate and star formation. Their results indicate that SFR and X-ray luminosity are mostly decoupled, with a correlation only at $z <1$ in galaxies with powerful AGN. This correlation disappears at higher redshifts, consistent with \cite{mullaney2012goods} and \cite{rosario2012mean}. \cite{hickox2014black} state that star formation and accretion activity are linked over long timescales but in low to moderate luminosity AGN the underlying correlation is hidden due to the AGN variability. Thus, we may not observe any direct correlation between SFR and the \emph{instantaneous} AGN luminosity in flux-limited AGN samples.

In this paper, we investigate the correlation between the SFR and stellar mass of AGN host galaxies with the nuclear activity of their SMBHs, using a large sample of galaxies with spectroscopic redshifts from the PRIsm MUlti-object Survey, PRIMUS, \citep{coil2011prism, cool2013prism}.
We use X-ray data from \textit{Chandra} and \textit{XMM-Newton} surveys that cover  $\sim 3$ deg${^2}$ of the PRIMUS area to identify a large sample of moderate-luminosity ($10^{41}<L_\mathrm{X}<10^{44} $ erg s$^{-1}$) AGN within our galaxy sample. We use X-ray luminosity as the tracer of AGN activity and estimate SFRs and stellar masses by fitting the observed galaxy spectral energy distribution (SED) using UV and optical photometry of our sources. With this data, we are able to probe down to relatively low SFRs and robustly separate our sample into quiescent and star forming populations. We also measure the fraction of AGN with star forming versus quiescent host galaxies (compared to a stellar mass-matched galaxy samples) and quantify how this fraction is changing with redshift, which could potentially drive any observed correlations in the overall sample. Finally, we measure the probability of a galaxy hosting an AGN and the distribution of specific accretion rates for both the star forming and quiescent galaxy populations, updating the study from \citet{aird2012primus} using our more robust galaxy classifications. We also further sub-divide the galaxy population to study the specific accretion rate distribution as a function of the specific SFR.

Section \ref{sec:data} briefly describes our data and the stellar mass completeness limits that we use to minimize observational biases. In Section \ref{sec:results} we describe our results on the correlation between SFR and AGN X-ray luminosity in our full sample.  We also investigate this correlation within sub-populations of star forming and quiescent galaxies. In addition, we consider the connection between galaxy star formation and AGN specific accretion rate and quantify the probability of a galaxy hosting an AGN as a function of specific SFR. We interpret our results in Section \ref{sec:discussion} and investigate the variation of the average X-ray luminosity of AGN with the star formation activity of their host galaxies. We summarize our results in Section \ref{sec:summary}. Throughout the paper we adopt a flat cosmology with $\Omega_{\Lambda}$ =0.7 and $H_0$=72 km s$^{-1}$Mpc$^{-1}$ and all magnitudes are on AB system.

\section{Data}
\label{sec:data}

In this study, we use multi-wavelength data from the PRIMUS survey covering four fields on the sky, including
 the \textit{Chandra} Deep Field South (CDFS), COSMOS, ELAIS S1, and XMM-LSS fields. 
All of these fields have deep UV, optical, and IR imaging as well as spectroscopic redshifts from PRIMUS. 
We also use X-ray imaging from \textit{Chandra} and \textit{XMM-Newton} to identify AGNs within the PRIMUS samples. 
We describe these datasets below, as well as our method of estimating stellar masses and SFRs
for our sources by fitting their spectral energy distributions (SEDs).

\subsection{PRIMUS}
\label{sec:primus}

We use data from PRIMUS, the largest faint galaxy, intermediate-redshift survey completed to date. The PRIMUS survey used the IMACS spectrograph on Magellan I Baade 6.5 m telescope at Las Campanas observatory, with a slitmask and a low-dispersion prism. 
The survey has a spectroscopic resolution of R $\sim$ 40 and covers a total of 9.1 deg${^2}$ of sky, spread over seven extragalactic 
fields with deep multiwavelength data.  
Objects were targeted to $i  \sim$ 23 using well-understood targeting weights. 
Four additional fields were targeted that had a large number of prior, high-resolution spectroscopic redshifts; these data were used for calibration purposes. 
In these calibration fields higher priority was given to targets with prior spectroscopic redshifts.  
The full details of the survey, targeting and data summary are presented in \cite{coil2011prism}.

In PRIMUS, objects are classified as stars, broad-line AGNs (BLAGNs) and galaxies based on their spectra. 
In each class, low-resolution spectra and multi-wavelength photometry of the objects are simultaneously fit with an empirical library of 
templates. For this study we restrict our sample to the sources with robust redshifts \citep[$Q\geq 3$, see ][]{coil2011prism}. The total PRIMUS catalog contains $\sim 120,000$ robust redshifts at $z\sim 0-1.2$, with a redshift precision of $\sigma_z/(1+z)\sim 0.005$. 
For further details of the data reduction, survey completeness, redshift fitting and precision see \cite{cool2013prism}.

PRIMUS targeted fields with existing deep multi-wavelength imaging. These data include X-ray imaging from \textit{Chandra} and \textit{XMM-Newton}, UV imaging from GALEX, deep optical imaging from a range of telescopes, and infrared imaging from the Infrared Array Camera (IRAC) and the Multiband Imaging Photometer (MIPS) on \textit{Spitzer}. In this paper, we restrict our analysis to spectroscopic sources targeted within the area with joint GALEX UV, optical, and  \textit{Spitzer} IRAC imaging. We thus restrict our sample to the COSMOS, ELAIS-S1 and XMM-LSS science fields in PRIMUS. In Sections \ref{sec:sfr_and_lx}--\ref{sec:sfr_and_lambda} we also use the PRIMUS calibration field CDFS-CALIB (hereafter CDFS), which overlaps with the deep \textit{Chandra} X-ray coverage. Taken together, these four fields cover 4.96 deg${^2}$ of the sky. We exclude the X-ray AGN and galaxy samples in the CDFS field below in the analysis of Section \ref{sec:pledd}, as it does not provide a uniformly targeted sample of galaxies, which is required for the analysis in that section.

\subsection{X-ray data}
\label{xray}

We use X-ray data to identify AGN within the PRIMUS galaxy sample. We have compiled X-ray catalogs in the CDFS, COSMOS, ELAIS S1, and XMM-LSS fields based on published \textit{Chandra} and \textit{XMM-Newton} surveys.
In the CDFS field, we use the \cite {lehmer2005extended} and \cite{luo2008chandra} X-ray catalogs corresponding to the 2 Ms observations of the central region (reaching depths of $f_\mathrm{2-8 keV} \sim 5.5 \times 10^{-17}$ erg s$^{-1}$ cm$^{-2}$) and the flanking 250 ks observations (reaching depths of $f_\mathrm{2-8 keV} \sim 6.7 \times 10^{-17}$ erg s$^{-1}$ cm$^{-2}$). The entire COSMOS field was observed with \textit{XMM-Newton} to depths of $f_\mathrm{2-10 keV} \sim 3 \times 10^{-15}$ erg s$^{-1}$ cm$^{-2}$ \citep{hasinger2007xmm}; additionally the central $\sim0.9$ deg$^2$ was observed with \textit{Chandra} to depths of $f_\mathrm{2-10 keV} \sim 8 \times 10^{-16}$ erg s$^{-1}$ cm$^{-2}$ \citep{elvis2009chandra}.
In the ELAIS-S1 field we use the catalog of \cite{puccetti2006xmm}, based on the \textit{XMM-Newton} observations that reach $f_\mathrm{2-10 keV} \sim 3 \times 10^{-15}$ erg s$^{-1}$ cm$^{-2}$.
Finally, in XMM-LSS we use X-ray data available from the both the \cite{pierre2007xmm} catalog and from the \textit{XMM-Newton} Deep Survey \citep{ueda2008subaru} down to the $f_\mathrm{2-10 keV} \sim 2 \times 10^{-15}$ erg s$^{-1}$ cm$^{-2}$. We use the likelihood ratio technique \citep[e.g.][]{sutherland1992likelihood,ciliegi2002deep,brusa2007xmm,laird2009aegis} to identify reliable optical counterparts (in the $i$ or $R$ band) to the X-ray sources; for objects with multiple counterparts, the match with the highest ratio is chosen. For full details of the construction of our X-ray catalogs and matching procedure see \cite{aird2012primus} and \cite{mendez2013primus}.

In this study, we consider a sample of obscured AGN detected in the hard (2--10 keV) X-ray band, where we have excluded objects that were classified as BLAGN in the PRIMUS spectra. These BLAGN constitute about 12$\%$ of our AGN sample.  As their optical emission can dominate over the optical light of the host galaxy, for such sources we are unable to estimate the stellar mass and SFR of the host. Restricting to hard X-ray (2--10 keV) detections ensures that we can estimate the X-ray luminosity with reasonable accuracy and are not strongly biased against the selection of moderately obscured sources.
Although hard X-ray emission passes through regions with moderate hydrogen column densities, it can not penetrate Compton-thick regions with heavy obscuration; thus our sample lacks this potentially important population.

We also restrict our analysis to sources with moderate X-ray luminosities in the range $10^{41}<L_\mathrm{X}<10^{44} $ erg s$^{-1}$ resulting in a final sample of 309 AGN. The lower limit of $10^{41}$ erg s$^{-1}$  ensures that the observed X-ray luminosity is dominated by light from an AGN rather than from star formation activity in the host galaxy. The upper limit ensures that the optical light of the host is not strongly contaminated by the presence of an AGN, such that we can estimate stellar masses and SFRs. It also ensures that our sample is not strongly biased by our exclusion of BLAGN, which constitute a higher fraction of the X-ray selected AGN population at high luminosities. To summarize, our AGN sample consists of 309 sources with hard-band X-ray detections and robust redshifts from PRIMUS (in the range $0.2<z<1.2$) that are not classified as BLAGN (thus the host galaxy dominates the optical light) and have X-ray luminosities within the range $10^{41}<L_X<10^{44} $ erg s$^{-1}$.

\subsection{Stellar Mass and SFR Estimates}
\label{sec:mass_and_sfr}

SED fitting is a widely adopted method for estimating the physical properties of galaxies. 
We estimate stellar masses and SFRs of the galaxies by fitting the observed SED based on the UV and optical photometry of our sources.  As we exclude
BLAGN and do not include IRAC photometry in our SED fits, we do not include an AGN contribution in the SED fitting.
We fit the SEDs using the \texttt{iSEDfit} code \citep{moustakas2013primus}, which is a Bayesian fitting code that compares the observed photometry for each source to a large Monte Carlo grid of SED models which span a wide range of stellar population parameters (e.g. age, metallicity, dust, and star formation history) to estimate the stellar mass and SFR of a galaxy.

\begin{figure}[ht!]
\includegraphics[width=0.5\textwidth]{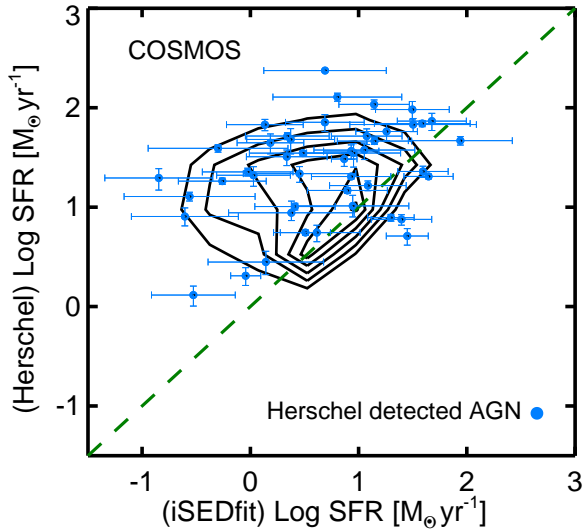}
\caption{
The SFR derived from \textit{Herschel} versus from \texttt{iSEDfit} for sources in the 
COSMOS field. We use \textit {Herschel} deep 100 $\mu$m observations and convert the FIR luminosity to a SFR using Equation \eqref{eq:lir}. Contours show the distribution of PRIMUS galaxies detected by \textit{Herschel}, while filled blue circles indicate AGN detected by \textit{Herschel}. The bulk of the AGN sample is not \textit{Herschel}-detected. The green dashed line indicates the 1:1 relation.
While the majority of \textit{Herschel}-detected galaxies lie close to this 1:1 line, there is a clear population that scatters above the line indicating that we underestimate the SFR using \texttt{iSEDfit}, although the shape of this distribution will be due to \textit{Herschel} only detecting dusty star-forming galaxies with high SFRs.
The \textit{Herschel}-detected X-ray sources span a similar space to the galaxies; upper limits on the \textit{Herschel} SFRs for X-ray sources without \textit{Herschel} detections place them in a similar space, confirming that the shape of the distribution is primarily driven by the limited depths of the \textit{Herschel} data.
}
\label{fig:herschel_isedfit}
\end{figure}

With \texttt{iSEDfit} we find the posterior probability distribution of stellar mass and SFR of a galaxy by marginalizing over all of the other parameters. We then take the median of the probability distribution functions as the best estimate of the stellar mass or SFR of each galaxy. 
The uncertainty on each parameter is calculated as one quarter of the 2.3--97.7 percentile range of the probability distributions, which would be equivalent to a $1\sigma$ uncertainty in the case of a Gaussian distribution. For details on \texttt{iSEDfit} see \cite{moustakas2013primus}.

For the SED fitting used in this paper, we adopt the Flexible Stellar Population Synthesis (FSPS) models \citep{conroy2009propagation,conroy2010propagation} with \cite{chabrier2003galactic} initial mass function (IMF) from 0.1 to 100 $\mathcal{M}_\odot $ and stellar metallicity in the range of $0.004<Z<0.05$. We consider exponentially declining star formation histories  $\Psi \propto \frac{1}{\tau}$ exp $(\frac{-t}{\tau})$, allowing for $\tau$ within the range of $0.01<\tau<1$ Gyr. We also allow for stochastic bursts of star formation on top of the smoothly decaying star formation histories. In addition, we include \cite{charlot2000simple} time dependent dust attenuation where attenuation in stellar population older than 10 Myr is less than younger populations \citep{charlot2000simple,wild2011empirical}. 

In this work, we estimate stellar masses and SFRs using \texttt{iSEDfit} on our 
UV and optical photometry. 
To test the accuracy of our SFRs, we compare them with \textit{Herschel} Space Observatory deep far infrared observations of the COSMOS field. We use deep 100 $\mu$m observations of the PACS Evolutionary Probe, PEP12  \citep{lutz2011pacs}, reaching a 3$\sigma$ limit of 5 mJy at 100 $\mu$m \citep{berta2011building}. We then use the \cite{kennicutt1998star} relation, given here as Equation \eqref{eq:lir}, using  Chabrier IMF to convert the FIR luminosity to SFR: 

\begin{equation} \label{eq:lir}
\frac{SFR}{\mathcal{M}_\odot yr^{-1}}= 1.09\times 10^{-10} \frac{L_{IR}}{L_{\odot}}  
\end{equation}

Figure \ref{fig:herschel_isedfit} compares the \textit{Herschel} derived SFR with our estimate from \texttt{iSEDfit} for PRIMUS sources in the COSMOS field. Contours show the distribution of PRIMUS galaxies that are detected by \textit{Herschel}, while blue circles show PRIMUS X-ray AGN that are detected by  \textit{Herschel}. The dashed line represents the 1:1 relation. As can be seen in the figure, the error bars on the \textit{Herschel} estimated SFRs are much smaller than those from \texttt{iSEDfit} and are likely underestimated, as a single template is used to calculate the total IR luminosity.

While many AGN and galaxies lie near the 1:1 relation, there is a sizable portion of the sample
well above the line, with the SFR estimated from \textit{Herschel} much higher than the SFR from 
\texttt{iSEDfit}. This is not surprising, as the 
IR luminosity is a more accurate probe of the SFR in dusty galaxies, while the inferred SFR from fitting the UV and 
optical SED primarily reflects unobscured star formation (though \texttt{iSEDfit} does fit and account for dust obscuration). 
Most PRIMUS X-ray AGN in COSMOS are {\it not} detected by \textit{Herschel}, and for these sources we calculate the 3$\sigma$ upper limits on SFR as estimated from the \textit{Herschel} imaging. These AGN not detected by \textit{Herschel} show a similar overall offset as the 
detected AGN in this figure, though we only have upper limits, such that the true values may lie close to the 1:1 line.  A histogram of the \textit{Herschel} to \texttt{iSEDfit} SFR differences in the \textit{Herschel}-detected galaxy sample 
peaks at $\Delta(\log SFR)=0$ but has a median offset of 0.6 dex. 
Within this sample, 42\% of the sources have a \textit{Herschel} SFR that is more than a
factor of three higher than the \texttt{iSEDfit} SFR,
although the shape of this distribution will be strongly skewed due to the limited depths of the \textit{Herschel} data. 

We note that, based on the KS test, the distributions of $\Delta(\log SFR)$ for the \textit{Herschel}-detected AGN and \textit{Herschel}-detected galaxies are not significantly different. Thus, the overall SFR estimated by \texttt{iSEDfit} for dusty galaxies may be systematically low, which appears to be due to \texttt{iSEDfit} underestimating the dust extinction in some of the \textit{Herschel}-detected galaxies (and thus underestimating the SFR). However, what we are interested in here is whether there is a correlation between SFR 
and $L_X$. A systematic offset will not affect our results, although additional scatter in our SFR estimates could wash out any underlying correlations.
Additionally, as \textit{Herschel} detects warm dust heated by star formation, the \textit{Herschel}-detected sample includes 
only the most dusty star forming galaxies, such that the SFR differences in the full PRIMUS galaxy and AGN sample will be less pronounced.

Below in Section \ref{sec:sf_and_q} we split the PRIMUS sample into star forming versus quiescent galaxies using the \texttt{iSEDfit} stellar mass and SFR values of each source. Here we estimate the contamination of our quiescent sample by star forming galaxies, by finding that in the COSMOS field 7\% of our quiescent sample (defined using \texttt{iSEDfit} outputs) is detected by \textit{Herschel}; these galaxies are therefore star forming galaxies and are misclassified.

\begin{figure*}[ht!]
\includegraphics[width=1.\textwidth]{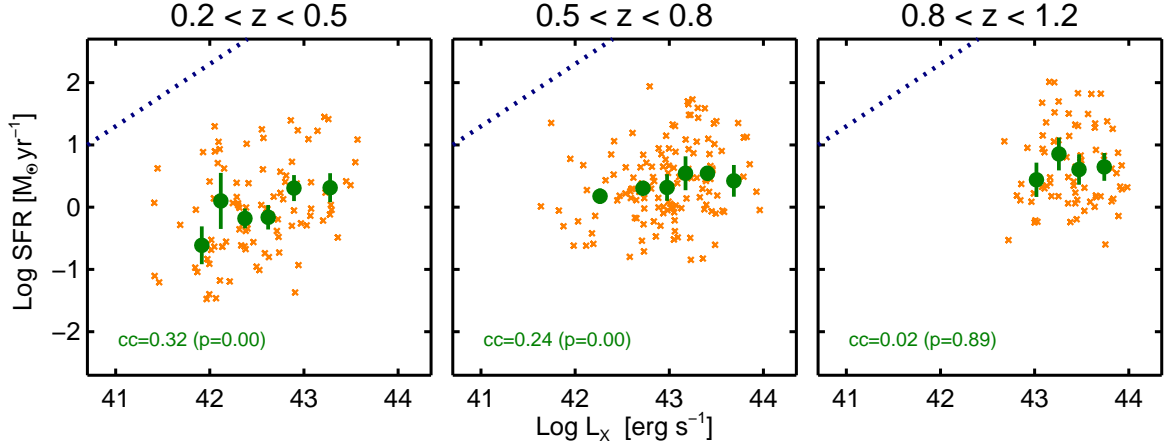}
\caption{{The SFR versus $L_X$ for a sample of non-broad line AGN in four PRIMUS fields, including CDFS, COSMOS, ELAIS S1 and XMM-LSS, for three redshift bins spanning $0.2 < z < 1.2$. 
Orange crosses show individual AGN, while green circles illustrate the median SFR in bins of $L_X$. The error bars show the uncertainty on our calculation of the median points, measured from bootstrap resampling.
The blue dotted line shows the SFR expected if the X-ray emission is from HMXBs; the fact that our sources are all well below this line indicates that the X-ray emission is from AGN. The correlation coefficients and correlation significance of the individual points in each panel are calculated from Spearman's rank correlation. A small $p$ value denotes it is unlikely for a correlation to have been occurred by accident and the correlation is considered significant if the $p$ value is less than 0.05. The lower two redshift panels show a weak trend between 
SFR and $L_X$, which is not apparent in the highest redshift panel, where our data probe a narrower range of $L_X$.
}}
\label{fig:lx_sfr_total}
\end{figure*}

\subsection{Stellar mass completeness limits}
 \label{sec:masscomp}

As PRIMUS is a flux-limited survey, targeting objects to $i\sim$ 23, this introduces a bias into our sample where we are unable to detect low-mass galaxies at higher redshifts, unless they have high SFRs (increasing the amount of blue light from the galaxy). To minimize this bias we define a stellar mass limit above which we can detect all galaxies, regardless of their SFR. This stellar mass limit is a smooth function of redshift and is slightly different in each field, depending on the band used for target selection  \citep[see also][] {aird2012primus}. Briefly, we define a template for a maximally old simple stellar population at $z \sim 5$ and calculate the mass-to-light ratio as a function of redshift, allowing the template population to evolve passively with time.
Then in each field we convert the targeting magnitude limit to a stellar mass limit as a function of redshift, as we keep only those galaxies with stellar masses above
this limit.  This restricts our sample to include only more massive galaxies at higher redshifts, but it ensures that we are not biased towards the star forming galaxy population and that we have a sample that is complete to a given stellar mass at all redshifts.  After applying these stellar mass limits, our final sample across the four 
fields used here consists of 32,865 galaxies at $0.2<z<1.2$, of which 283 are AGN detected in the hard X-ray band with $10^{41}<L_\mathrm{X}<10^{44}$ erg s$^{-1}$.

\section{Results}
\label{sec:results}

In this section, we investigate the relationship between SFR and X-ray 
luminosity, $L_\mathrm{X}$, in our AGN sample. In order to uncover whether 
stellar mass could be an underlying variable, we also investigate 
the stellar mass dependence of $L_\mathrm{X}$. We further divide our full
AGN sample into those with star forming and quiescent host galaxies,
to consider how SFR and stellar mass vary with $L_\mathrm{X}$ within each host 
population.  We also measure the probability that a galaxy of a given 
stellar mass and redshift hosts an AGN, as a function of the specific SFR
of the galaxy.

\begin{figure*}[ht!]
\includegraphics[width=1.\textwidth]{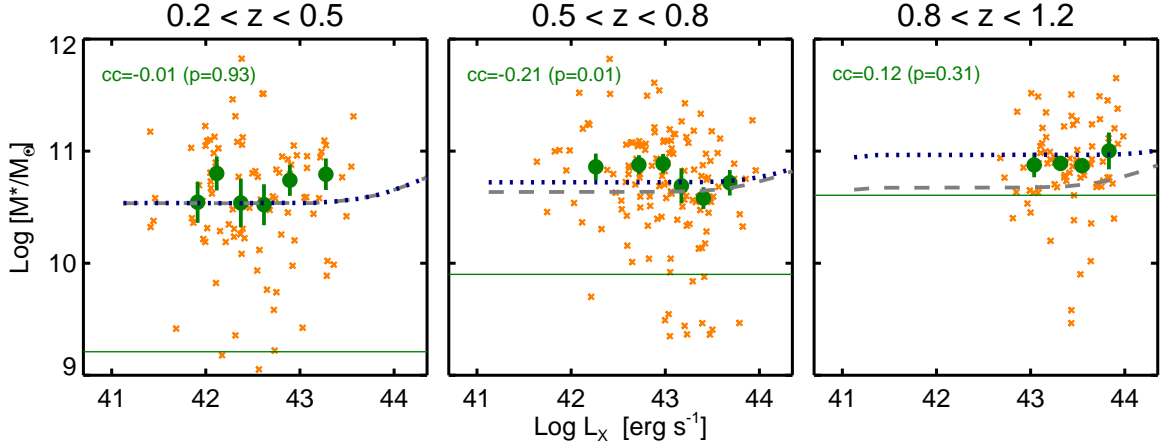}
\caption{{Stellar mass versus $L_\mathrm{X}$ for our sample of non-broad AGN in three redshift bins spanning $0.2 < z < 1.2$. Orange crosses show individual AGN while green circles illustrate the median stellar mass in bins of $L_\mathrm{X}$. The error bars show the uncertainty on median points, measured from bootstrap resampling. The solid dark green line indicates the PRIMUS stellar mass completeness limit. The grey dashed line is a prediction from a model presented in \cite{aird2013primus} and shows the predicted median stellar mass as a function of $L_\mathrm{X}$ in each redshift range. The blue dotted line shows the prediction of this model for sources above the PRIMUS mass completeness limit. There is no significant trend between stellar mass and $L_\mathrm{X}$  in the first and last panel but there is a negative correlation in the middle panel that is mainly due the sources below the mass completeness limit.}}
\label{fig:mass_lx_total}
\end{figure*}

\subsection {The relationship between SFR and stellar mass with $L_\mathrm{X}$}
\label{sec:sfr_and_lx}

Figure \ref{fig:lx_sfr_total} shows the SFR of AGN host galaxies plotted as a function of $L_\mathrm{X}$, in three redshift bins spanning $0.2 < z < 1.2$.  Orange crosses show individual AGN, while green circles show the median SFR in bins of $L_\mathrm{X}$, where each bin contains at least 15 sources, to visually highlight any correlations. 
X-ray emission at these luminosities can arise not only from AGN but potentially from high mass X-ray binaries (HMXBs), which are tracers of star formation 
activity and generally have lower luminosities, $L_\mathrm{X}\sim 10^{35−-40}$ erg s$^{-1}$.  The dotted blue line in Figure \ref{fig:lx_sfr_total} is from \cite{ranalli20022} and shows the relation between SFR and $L_\mathrm{X}$ for HMXB: 
\begin{equation} \label{eq:ranalli}
\frac{SFR}{\mathcal{M}_\odot yr^{-1}}=2.0 \times 10^{-40} L_\mathrm{X (2-10  \  Kev)} \  erg \;s^{-1}                         
\end{equation}
As this line is well above our sources, it indicates that the
X-ray emission seen for our sample is from AGN and does not suffer from contamination by HMXBs.

For the median points shown in Figure \ref{fig:lx_sfr_total}, we estimate error bars using 
bootstrap resampling. The uncertainty shown reflects the variance among 
the median SFR in each of 1000 bootstrap samples.
These errors are similar to the standard errors calculated in each $L_\mathrm{X}$ bin. 
This figure clearly shows that there is a wide spread in SFR at any given value of $L_\mathrm{X}$, such that the standard deviation of the points is typically 3--4 times greater than the error shown.

We use the $r$\_$correlate$ routine in IDL to find the correlation coefficients and correlation significance of the individual points in Figure \ref{fig:lx_sfr_total}. This routine computes the Spearman's rank correlation coefficient and the significance of its deviation from zero,  $p$. This value indicates the probability of obtaining a desired event under the null hypothesis that the event happened purely by chance. A small $p$ value denotes it is unlikely for the correlation to have been occurred by accident. A correlation is considered significant if the $p$ value is less than 0.05. We quote $p$ values to an accuracy of two decimal places, thus $p=0.00$ indicates cases where we can reject the null hypothesis at a confidence level of $>$99.5\%. The correlation coefficients for the individual points shown in Figure \ref{fig:lx_sfr_total} are 0.32 ($p$=0.00), 0.24 ($p$=0.00), and 0.02 ($p$=0.89), respectively, from the lowest to the highest redshift bin.

In this figure, there is a large scatter in SFR in bins of $L_\mathrm{X}$ in all three redshift ranges.  At a given $L_\mathrm{X}$, the average
SFR increases at higher redshifts, consistent with the overall increase in SFR seen in the galaxy population \citep[e.g.][]{Bell05,elbaz2007reversal,noeske2007star}.
In the two lowest redshift ranges we see a weak positive correlation in the median points that is confirmed by the significance of the correlation coefficients measured above. We find  no correlation between SFR and $L_\mathrm{X}$ in the highest redshift range, but we note that at higher redshifts we are probing a more limited range of X-ray luminosity. 

To determine whether the observed trends in the first two panels are actually being driven by redshift-dependent selection effects, we measure the median redshift in the each bin of $L_\mathrm{X}$. As the median redshift does not systematically vary with $L_\mathrm{X}$, redshift is not driving any trends in Figure \ref{fig:lx_sfr_total}.

However, the weak correlation between SFR and $L_\mathrm{X}$ seen in the lower two redshift panels could be due to 
an underlying trend between stellar mass and $L_\mathrm{X}$.
Given that within the galaxy population there is a positive correlation between SFR and stellar mass  \citep[e.g.][]{elbaz2007reversal,karim2011star}, it is possible that the observed correlation between SFR and $L_\mathrm{X}$ could
actually be due to an underlying correlation between stellar mass and $L_\mathrm{X}$.

In Figure \ref{fig:mass_lx_total} we show the stellar mass of AGN host galaxies as a function of $L_\mathrm{X}$. As in Figure \ref{fig:lx_sfr_total}, orange crosses indicate individual sources while green circles show median values in bins of $L_\mathrm{X}$, and errors on the median points are calculated using bootstrap resampling. The standard deviation of the data points is larger than the error shown by a factor of 3--5. 
The correlation coefficients of the individual points in this figure are -0.01 ($p$=0.93), -0.21 ($p$=0.01) and 0.12 ($p$=0.31), respectively, for the three
redshift ranges shown. 
We note that we have not applied the stellar mass completeness limits discussed above in Section \ref{sec:masscomp} and we show our full X-ray AGN sample in both Figures \ref{fig:lx_sfr_total} and \ref{fig:mass_lx_total}.
In Figure \ref{fig:mass_lx_total}, we only find a significant correlation in the middle redshift range; however, the correlation is \emph{negative} and appears to be driven by a small number of sources below the PRIMUS mass completeness limit shown with the solid dark green line. 
If we only consider sources above the stellar mass completeness limits, we do not find any significant correlation between stellar mass and $L_\mathrm{X}$ for X-ray AGN in any of the three redshift ranges.
We thus conclude that the observed (weak, but significant) positive correlations between SFR and $L_\mathrm{X}$ in Figure \ref{fig:lx_sfr_total} are \emph{not} due to a positive correlation between stellar mass and $L_\mathrm{X}$ within our X-ray AGN sample.

The grey dashed line in Figure \ref{fig:mass_lx_total} is a prediction from a model presented in \cite{aird2013primus}. This model takes the stellar mass function of galaxies and populates the galaxies with AGN using a universal power-law distribution of specific accretion rates (the rate of accretion scaled relative the host stellar mass, see Section \ref{sec:fraction} below). The specific accretion rate distribution itself does not depend on stellar mass but has a normalization that evolves with redshift  \citep[motivated by the observational results of][] {aird2012primus}. The grey dashed line shows the predicted median stellar mass as a function of
$L_\mathrm{X}$, using this model, and in particular shows that the median stellar mass of AGN
host galaxies should not vary significantly with $L_\mathrm{X}$, over the luminosity 
range where we have data. The blue dotted line shows the prediction of this model for sources above the PRIMUS mass completeness limits, and the median points in our sample lie close to this line, confirming this lack of a correlation.
 
\begin{figure*}[ht!]
\centering
\includegraphics[width=1.\textwidth]{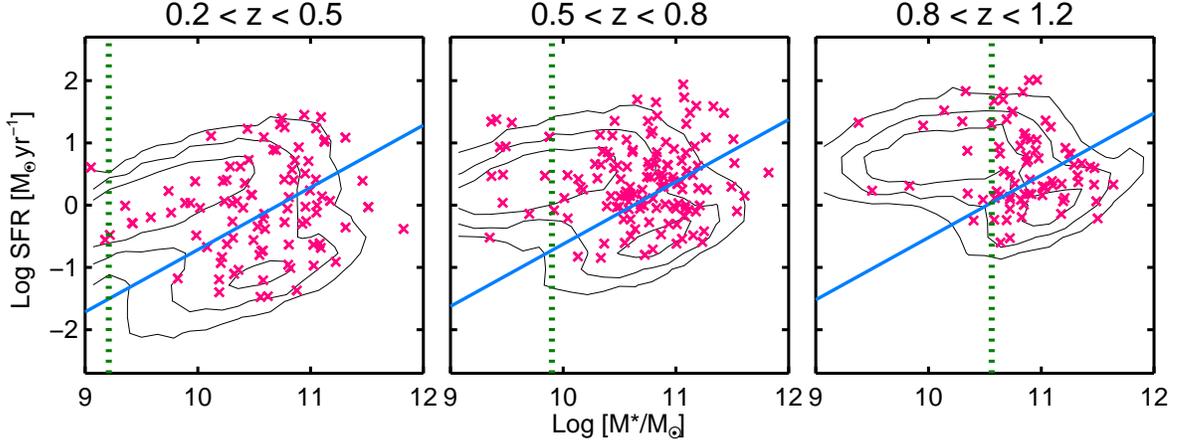}
\caption{SFR versus stellar mass for PRIMUS galaxies and AGN, shown in three redshift ranges. Contours show the distribution of galaxies in this space; there are two distinct
populations: star forming galaxies, with a relatively high SFR at a given stellar mass, and quiescent galaxies, with a low SFR at a given stellar mass.  Red crosses show X-ray AGN, which reside in both star forming and quiescent host galaxies. The blue solid line is the classification used to define galaxies as being either star forming (above the line) or quiescent (below the line). This line evolves with redshift. The vertical dotted green line shows the PRIMUS stellar mass completeness limit, which is also a function of redshift.  To create samples that are complete in stellar mass, we exclude sources to the left of this line.
}
\label{fig:sfr_mass}
\end{figure*}

There is a large scatter in the stellar mass at any given X-ray luminosity. 
We find that the average stellar mass of the AGN host galaxies in all three redshift ranges is higher than $10^{10} {\mathcal{M}_\odot }$. This is consistent with prior literature, including   \cite{kauffmann2003host,xue2010color} and \cite{aird2012primus}, who find that observed AGN are predominantly hosted by moderately massive galaxies. 
While some recent studies have found evidence for AGN activity in much lower mass, dwarf galaxies, such sources represent a small fraction of the X-ray selected population and have X-ray luminosities below our limit \citep[e.g.][]{Moran99,Barth05,Reines11,reines13,Secrest}.
Nonetheless, we note that \cite{aird2013primus} attribute the dominance of moderate-mass host galaxies in X-ray--selected AGN samples to the shape of the stellar mass function of galaxies, combined with the wide power-law distribution of AGN accretion rates, rather than an enhancement of AGN activity in galaxies of a particular stellar mass. 

Overall, Figure  \ref{fig:mass_lx_total} demonstrates that over the X-ray luminosity that we probe, $41< \log L_\mathrm{X} < 44$, there is no correlation between AGN luminosity and host stellar mass within the AGN sample (although we note that our AGN sample is dominated by moderately massive galaxies, $\mathcal{M}_*>10^{10} \mathcal{M}_\odot$ at all luminosities and redshifts). The weak correlation seen above between SFR and X-ray luminosity is therefore not due to an underlying correlation between $L_\mathrm{X}$ and stellar mass within the AGN sample.

\subsection {The relationship between SFR and stellar mass with $L_\mathrm{X}$ for star forming and quiescent galaxies}
\label{sec:sf_and_q}

Our results above indicate that 
there is a weak but significant correlation between SFR and $L_\mathrm{X}$ for AGN 
host galaxies for the lowest two redshift bins probed here, spanning $0.2 < z < 0.8$. We do not find a significant positive correlation between stellar mass and $L_\mathrm{X}$ for the same sample, indicating that stellar mass is not driving the observed
SFR-$L_\mathrm{X}$ relation. We further investigate possible effects of the host population by 
splitting the AGN host galaxies into star forming and quiescent populations 
and determining whether a SFR-$L_\mathrm{X}$ relation exists within either of 
these populations alone.
 
We classify each AGN host galaxy as star forming or quiescent using its 
specific star formation rate, $sSFR$, which is defined as the SFR per unit stellar mass, $\frac{SFR}{\mathcal{M}_*}$. Histograms of $ \log sSFR$ of our full galaxy sample within relatively narrow redshift ranges show two prominent peaks: one corresponding to the ``main sequence'' of star forming galaxies \citep{noeske2007star} and the other to quiescent galaxies. The locations of the peaks evolve with redshift, in that $sSFRs$ are higher (on average) at higher redshift.  To classify galaxies into 
star forming or quiescent, we wish to use the $sSFR$ corresponding to the minimum between these two populations in the sSFR histogram.  This minimum is not always clear in all of the (six) redshift bins used, but the peak of the star forming main sequence can be traced easily at all redshifts.  We therefore first fit for the evolution of this peak, $sSFR_{max}$, as a linear function of redshift, which is given by:
\begin{equation} \label{eq:ssfrmax}
\log (sSFR_{max}(z))=0.3\times z-9.62  
\end{equation}

We then normalize the $sSFR$ of each galaxy relative to this $sSFR_{max}(z)$, which we 
call the epoch normalized specific star formation rate, $ENsSFR$  (see also \cite {stott2012xmm})  :
\begin{equation} \label{eq:enssfr}
\log (ENsSFR)=\log (sSFR)-0.3\times z+9.62                            
\end{equation}

\begin{figure*}[ht!]
\centering
\includegraphics[width=1.\textwidth]{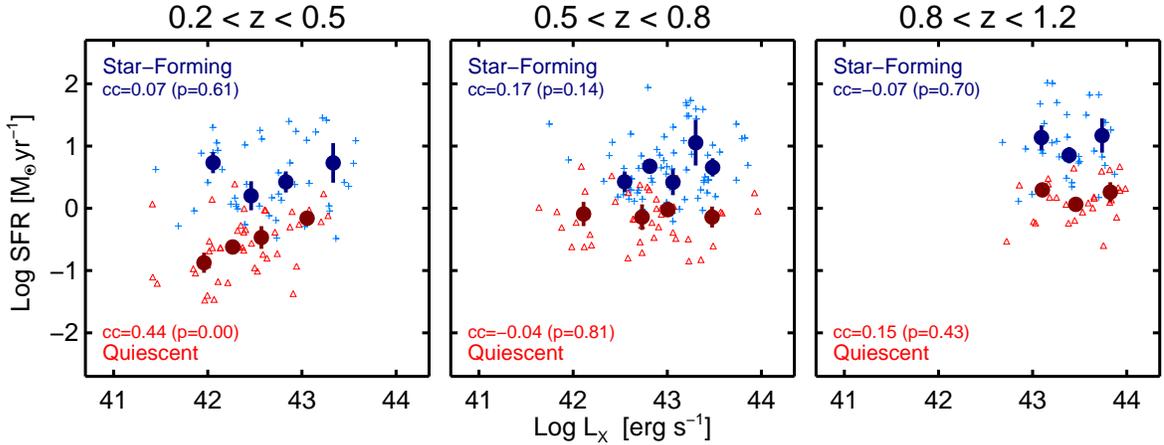}
\caption{The SFR versus $L_\mathrm{X}$ of AGN host galaxies, in three redshift bins, where the host galaxies are split into star forming (blue plus) or quiescent (red triangle).
 Similar to Figure \ref{fig:lx_sfr_total}, the errors are measured using the bootstrap resampling. 
Blue and red circles show the median SFR in bins of $L_\mathrm{X}$, for the star forming and 
quiescent host populations, respectively. There is a significant positive trend in quiescent host population in the lowest redshift panel that vanishes at higher redshifts. There is no significant correlation between SFR and  $L_\mathrm{X}$ in star forming galaxies in any redshift bins. }
\label{fig:sfr_lx}
\end{figure*}
\begin{figure*}[ht!]
\centering
\includegraphics[width=1.\textwidth]{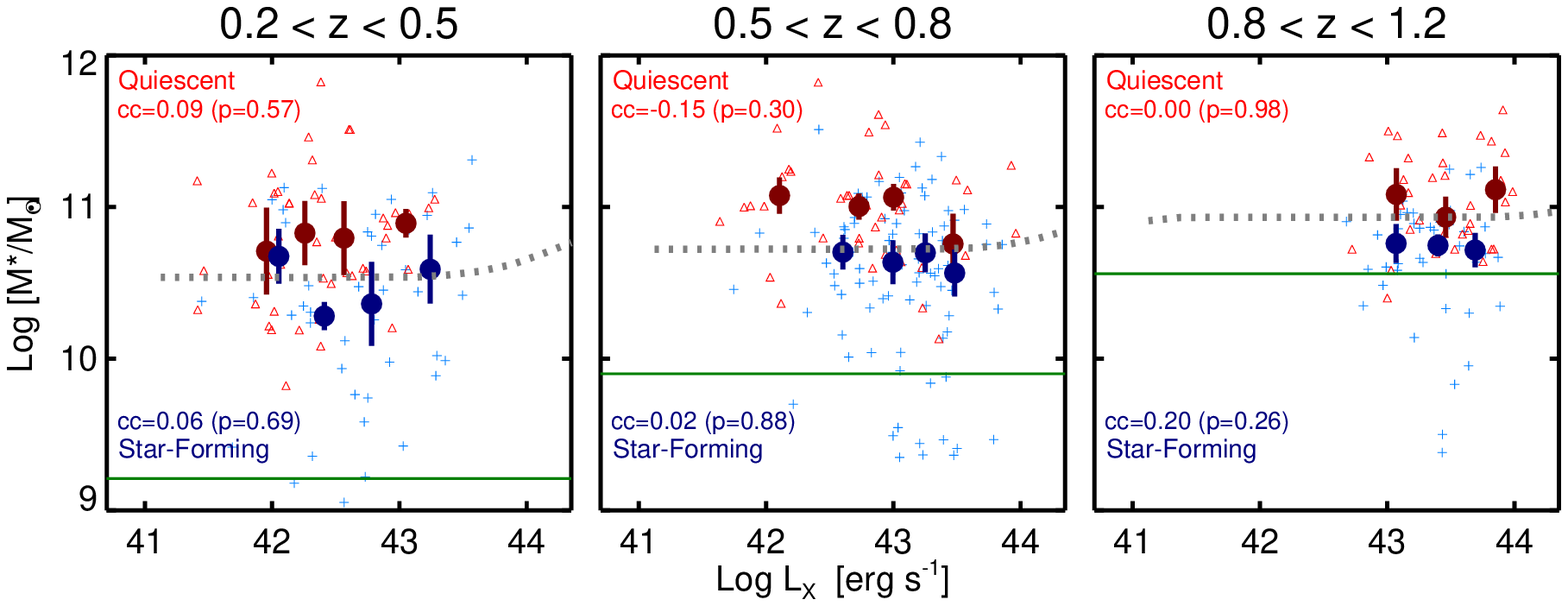}
\caption {The average stellar mass versus $L_\mathrm{X}$, in three redshift ranges. The host galaxies are split into star forming (blue plus) or quiescent (red triangle). The errors are measured using the bootstrap resampling. Similar to Figure \ref{fig:mass_lx_total}, the grey dashed line is a prediction of stellar mass as a function of  $L_\mathrm{X}$ from a model presented in \cite{aird2013primus} for sources above PRIMUS mass completeness limit. The green solid line shows the PRIMUS stellar mass completeness limit, which is also a function of redshift. There is no significant correlation between stellar mass and $L_\mathrm{X}$ in either of populations above the mass completeness limit.}
\label{fig:mass_lx}
\end{figure*}

Finally, we plot a histogram of $\log ENsSFR$ for all of galaxies, which exhibits a clear bimodality. 
We find the minimum between the two peaks of the bimodal distribution at $\log ENsSFR_{min}=-1.2$.
We divide our sample into star forming and quiescent galaxies according to whether their $ENsSFR$ is above or below $ \log ENsSFR_{min}$, respectively. Ultimately, we apply the same classification scheme to our X-ray AGN sample.

Figure \ref{fig:sfr_mass} shows the location of PRIMUS galaxies and AGN within the SFR-stellar mass plane, in the same three redshift ranges as used above. The contours indicate the location of PRIMUS galaxies, while red crosses show individual AGN host galaxies. The blue line shows the separation defined above between the star forming and quiescent populations (shown at the median redshift of each redshift range); galaxies above the line are considered to be star forming while galaxies below are classified as quiescent. The vertical dashed green line shows the stellar mass completeness limit for PRIMUS, at the median redshift of that panel.  Above this stellar mass limit we are complete for both star forming and quiescent galaxies (see Section \ref{sec:masscomp} above). 
Figure \ref{fig:sfr_mass} shows that AGN are present in both the star forming and quiescent galaxy populations. However, the most massive galaxies (and therefore AGN host galaxies) in the sample tend to be quiescent, especially at $z>0.5$.

\begin{figure*}[ht!]
\centering
\includegraphics[width=1.\textwidth]{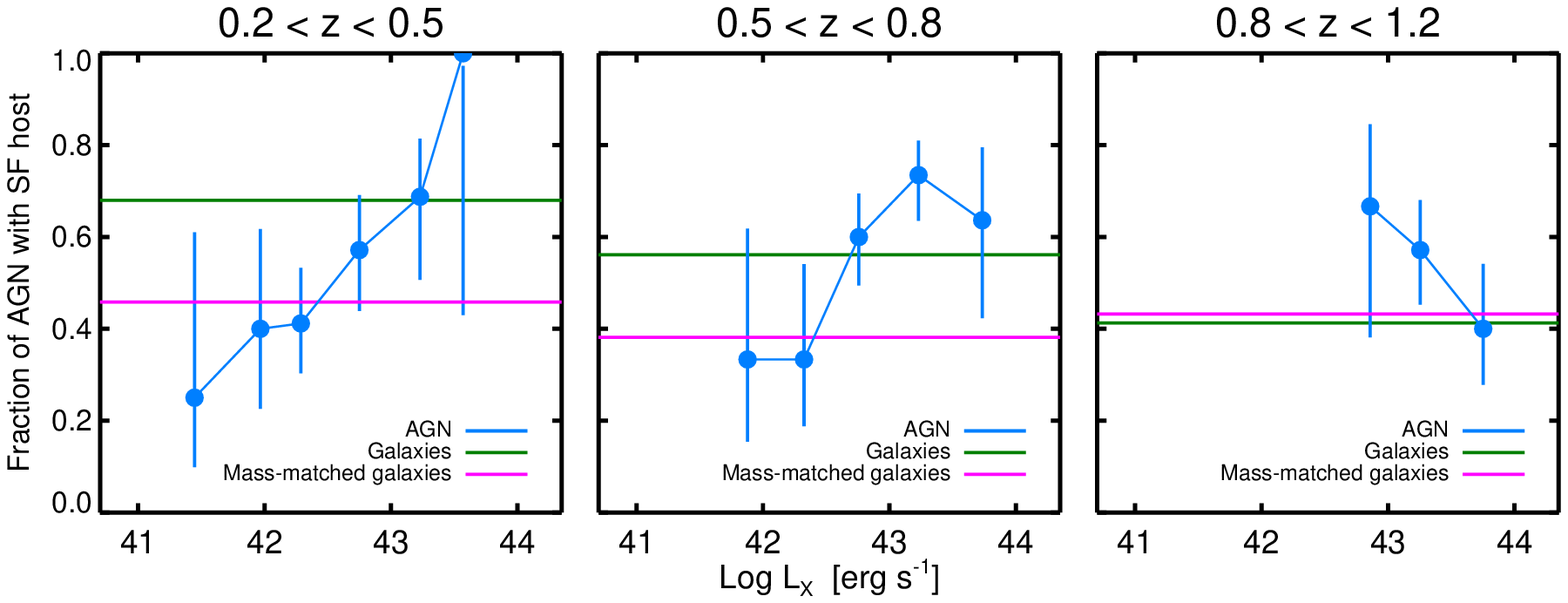}
\caption{Variation of the fraction of AGN in star forming host galaxies with $L_\mathrm{X}$, for stellar mass complete samples, shown with blue lines. 
The errors are calculated from the binomial distribution using the Bayesian method of \cite{cameron2011estimation}.
The green lines show the fraction of all galaxies above the stellar mass completeness limits that are star forming, while the pink lines show this fraction for galaxy samples
that have the same stellar mass distributions as the AGN host galaxies. While the star forming host fraction increases with increasing $L_\mathrm{X}$ in the lower two redshift ranges, given the error bars neither trend is significant. In the middle redshift range there is a 2$\sigma$ difference between the total fraction of AGN host galaxies that are star forming and the fraction of stellar mass matched galaxies that are star forming.}
\label{fig:fraction}
\end{figure*}

We apply the above star forming versus quiescent classification to all PRIMUS galaxies, and we apply the stellar mass completeness limit as well, and show the SFR versus $L_\mathrm{X}$ of AGN host galaxies for each populations in Figure \ref{fig:sfr_lx}.  The blue points show host galaxies on the star forming main sequence, while the red points show quiescent host galaxies. 
We find the median SFR in bins of X-ray luminosity, where we require a minimum of 12 and 10 AGN per bin for the star forming and quiescent populations, respectively. We note that these median points are for the purpose of illustrating the trends in data and the numbers are chosen to have at least three $L_\mathrm{X}$ bins for each population in each panel. As above, error bars on the median points are from bootstrap resampling and the standard deviation in each population is larger than the errors shown by a factor of 3--4. We find a fairly high scatter in the SFR at a given $L_\mathrm{X}$ within each host galaxy population. However, due to the flux limit of the survey we probe a smaller range of $L_\mathrm{X}$ in the highest redshift panel.

Within the star forming population, the lack of a trend in the median points and the correlation coefficients (none of which are significant, as seen in the figure) confirm the absence of any significant correlation between SFR and $L_\mathrm{X}$ in all three redshift ranges probed here.  Within the quiescent population there is a weak but significant trend in the lowest redshift range only; the correlation coefficients are 0.44 ($p$=0.00), -0.04 ($p$=0.81) and 0.15 ($p$=0.43), respectively from lowest to highest redshift.

We further show in Figure \ref{fig:mass_lx} the stellar mass versus $L_\mathrm{X}$ of the AGN host galaxies, split into the star forming and quiescent populations. As in Figure \ref{fig:sfr_lx}, the blue and red symbols represent star forming and quiescent galaxies, respectively, and the error bars are calculated using bootstrap resampling. The horizontal green solid line shows the PRIMUS stellar mass completeness limit at the median redshift of each panel, and the grey dotted line is from the model of \cite{aird2013primus}, as described in Section \ref{sec:sfr_and_lx}, for sources above this completeness limit. 

Both galaxy populations have a range of stellar masses but on average the stellar masses of the quiescent galaxies are higher in all three redshift ranges. Applying the stellar mass completeness limit clearly narrows the dynamic range of our sample in stellar mass at higher redshifts, particularly for the star forming population. As with the full galaxy population, we do not find any significant correlation between stellar mass and $L_\mathrm{X}$ for either the star forming or quiescent host galaxy populations in any of the three redshift ranges. We also find that applying the stellar mass completeness limit and splitting our sample to star forming and quiescent, the weak trend found above in Figure \ref{fig:mass_lx_total} for the middle redshift range now vanishes. 

Overall, it appears that the weak positive trend in the first panel of Figure  \ref{fig:lx_sfr_total} is due to a correlation between SFR and $L_\mathrm{X}$ in the quiescent host population. Figure \ref{fig:mass_lx} further shows that this trend is not due to the stellar mass.
For the other redshift ranges, we no longer find a significant correlation between SFR and $L_\mathrm{X}$ after after applying the stellar mass completeness limit and splitting the sample into the star forming and quiescent host galaxies (although we note that splitting up our sample, in itself, could eliminate any weakly significant trends seen in the full sample).

\subsection {The fraction of star forming AGN host galaxies}
\label{sec:fraction}

In Section \ref{sec:sfr_and_lx} we found that  in the two lowest redshift panels of Figure \ref{fig:lx_sfr_total} there is a weak but significant positive correlation between SFR and $L_\mathrm{X}$ in AGN host galaxies. In Section \ref{sec:sf_and_q} we demonstrated that when splitting the sample of these host galaxies into star forming and quiescent, this trend disappeared except for quiescent AGN hosts in the lowest redshift bin. We now investigate whether a change in the fraction of star forming host galaxies, as a function of $L_\mathrm{X}$, could be driving the observed correlation between SFR and $L_\mathrm{X}$ for the full AGN host sample. For example, a higher fraction of star forming host galaxies at the most luminous end in Figure  \ref{fig:lx_sfr_total} could create a positive trend in the full sample, as is seen.

\begin{figure*}[ht!]
\centering
\includegraphics[width=1.\textwidth]{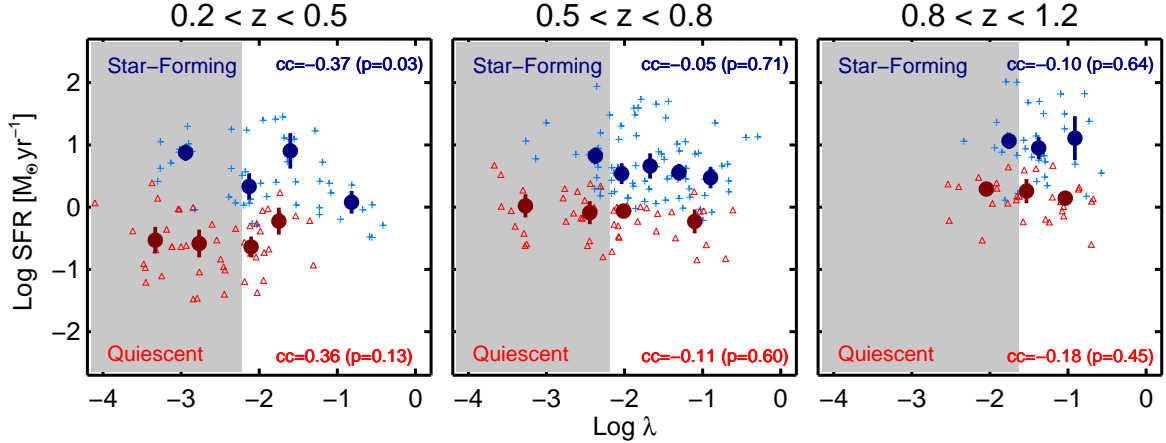}
\caption{The average SFR versus specific accretion rate, in three redshift bins, where the host galaxies are split into star forming (blue plus) or quiescent (red triangle). The errors are measured using bootstrap resampling. 
Blue and red circles show the median SFR in bins of $\lambda$, for the star forming and quiescent host populations, respectively. The grey regions show the area below the specific accretion rate limit; sources in these regions are excluded from the sample. There is no significant correlation between SFR and $\lambda$ within the quiescent host population. There is a 2$\sigma$ correlation in the star forming host population in the lowest redshift range that is not seen at higher redshifts.} 
\label{fig:sfr_lambda}
\end{figure*}

We calculate the fraction of X-ray selected AGN with a star forming host galaxy as a function of $L_\mathrm{X}$, where we consider only galaxies above the stellar mass completeness limits. 
The results are shown in Figure \ref{fig:fraction} in bins of 0.5 dex in $L_\mathrm{X}$, in three redshift ranges. The errors on the fractions are calculated assuming a binomial distribution using the Bayesian method of \cite{cameron2011estimation} and are equivalent to 1$\sigma$ uncertainties (68.3\% equal-tail confidence intervals).
To compare the AGN host galaxies with inactive galaxies in the same redshift range, 
we show the fraction of entire PRIMUS galaxies above the stellar mass completeness limit that are star forming with green solid lines. Although we find AGN in galaxies with a wide range of stellar mass, they are mainly found in relatively massive galaxies.
Furthermore, the median stellar mass of the AGN host galaxies increases from $10^{10.6}$ to $10^{10.8}$ and $10^{10.9} \; {\mathcal{M}_\odot }$ respectively over the three redshift ranges shown in Figure \ref{fig:fraction}, primarily due to the increasing stellar mass limit of the PRIMUS sample.
For each redshift range we therefore make a sample of stellar mass-matched galaxies that has the same stellar mass distribution as the AGN host galaxies. Then we compare the fraction of AGN within star forming host galaxies with a sample of inactive galaxies with a similar stellar mass and redshift distribution.
To construct this sample, we weight each galaxy in a redshift bin such that the  weighted distribution of stellar masses matches the stellar mass distribution of the X-ray AGN host galaxies.  The fraction of the stellar mass-matched galaxy sample that is star forming is shown with pink lines in Figure \ref{fig:fraction}.

\begin{figure*}[ht!]
\centering
\includegraphics[width=1.\textwidth]{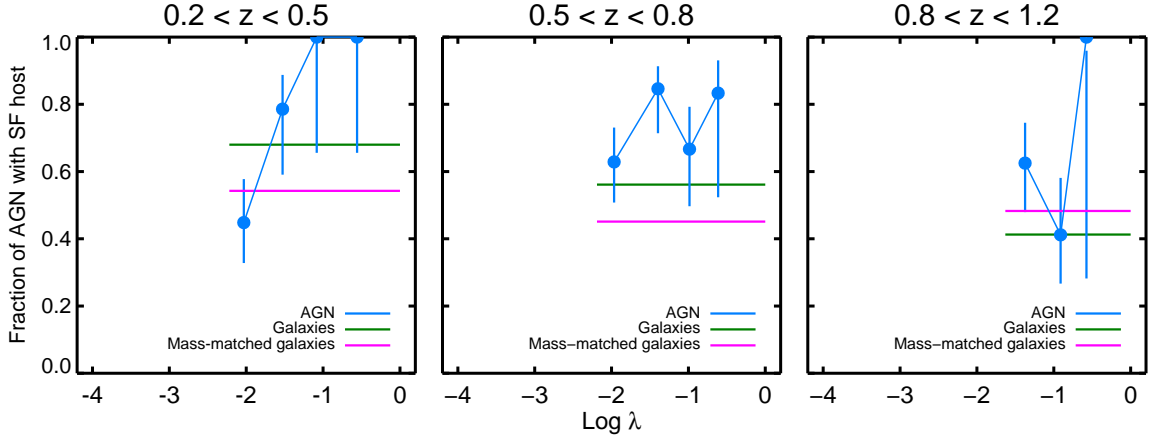}
\caption{Variation of the fraction of AGN in star forming hosts with $\lambda$, for the specific accretion rate complete sample, shown with blue lines. The errors are calculated from the binomial distribution using the Bayesian method of \cite{cameron2011estimation}. The green lines show the fraction of all galaxies above the specific accretion rate limits that are star forming, while the pink lines show this fraction for galaxy samples
that have the same stellar mass distributions as the AGN host galaxies. The fraction of star forming hosts increases with increasing $\lambda$ in the lowest redshift. The difference between the fraction of star forming hosts and mass-matched galaxy sample is less than 1$\sigma$ in the first and last panel and is less than 2$\sigma$ in the middle panel, confirming that these fractions are not significantly different.}
\label{fig:fraction_lambda}
\end{figure*}

In the lowest redshift range, $0.2<z<0.5$, there is a strong apparent trend such that the fraction of X-ray AGN in star forming hosts increases with increasing X-ray luminosities, from $\sim$20\% at $L_\mathrm{X} = 10^{41.5}$ erg s$^{-1}$ to 100\% at $L_\mathrm{X}=10^{43.5}$ erg s$^{-1}$. This trend is generally consistent with other studies \citep[e.g.][]{kauffmann2003host,heckman2006host} in a similar redshift regime that found that low luminosity AGN typically reside in early type galaxies with low star formation activity, while powerful AGN reside in galaxies with young stellar populations.  In the middle redshift range, $0.5<z<0.8$, the fraction also increases with increasing X-ray luminosity, though not as strongly as at lower redshift. However, in the highest redshift range, $0.8 < z < 1.2$, the fraction declines with X-ray luminosity, over the more limited range spanned by the data at these higher redshifts. 
However, considering the large error bars, none of these trends are significant.

In an attempt to decrease the error bars, in each redshift range we split the AGN sample into just two equal-width bins in $L_\mathrm{X}$. In the lowest redshift range, for 41$<\log L_\mathrm{X} <$42.5 this fraction is 40$\pm$19\% and increases to 63$\pm$20\%  at higher luminosities of 42.5$<\log L_\mathrm{X} < $44. In the middle redshift range, the fraction increaes from 33$\pm$30\% to 67$\pm$13\% in the two luminosity bins. Considering the error bars, neither of these increases is significant. 

\begin{table}
\caption{ The fraction of X-ray selected AGN with star forming host galaxies, and the fraction of stellar mass-matched galaxies that are star forming.}
\begin{center}
  \begin{tabular}{| c | c | c |}
    \hline
    Redshift& AGN with SF hosts & SF mass-matched galaxies \\
    \hline
    $0.2<z<0.5$ & 51$\pm$14\% & 46\% \\ \hline
    $0.5<z<0.8$ & 62$\pm$12\% & 38\% \\ \hline 
    $0.8<z<1.2$ & 52$\pm$17\% &  43\% \\
    \hline

  \end{tabular}
\end{center}
\label{tab:fraction}
\end{table}

Table \ref{tab:fraction} lists the overall fraction of AGN host galaxies that are star forming, as well as the fraction of stellar mass-matched galaxies that are star forming, in each redshift range. The variation of the star forming fraction for the stellar mass-matched galaxy sample with redshift is due to the combination of the PRIMUS stellar mass limits (which restrict us to higher stellar masses at higher redshifts), the preference for observed X-ray AGN to be found in more massive galaxies (across all redshifts), and the intrinsic changes in the star forming fraction for galaxies of a given stellar mass with increasing redshift  \citep[e.g.][]{moustakas2013primus}. The difference in the star forming fraction of AGN host galaxies compared to that of stellar mass-matched galaxies is less than 1$\sigma$ and therefore is not significant in the lowest and highest redshift bins; however, these fractions are different at the 2$\sigma$ level in the middle redshift range. The fraction of AGN with star forming host galaxies across our full redshift range of $0.2 < z < 1.2$ is 55$\pm$8\%, and the fraction of star forming galaxies in the corresponding mass-matched galaxy sample is 44\%. Therefore, across our full redshift range, there is not a significant difference in the the fraction of AGN hosted by star forming galaxies and the fraction of star forming galaxies with the same stellar mass distribution.

We find that the fraction of AGN with star forming host galaxies 
appears to increase with $L_\mathrm{X}$ in the two lowest redshift ranges, spanning $0.2<z<0.8$, but considering the large error bars the trends observed are not significant. However, the star forming host fraction does increase from $\sim30$\% to $\sim100$\% as $L_\mathrm{X}$ increases, at least in the lowest redshift range, such that the correlation between SFR and $L_\mathrm{X}$ observed at $0.2 < z < 0.8$ in Figure  \ref{fig:lx_sfr_total} could be due to the increasing fraction of star forming host galaxies with $L_\mathrm{X}$.

\subsection {The relationship between SFR and specific accretion rate}
\label{sec:sfr_and_lambda}

In Section \ref{sec:sf_and_q} we showed that there is a wide range in the stellar masses of AGN host galaxies for both star forming and quiescent populations. This large scatter could potentially hide an underlying correlation between SFR and  specific accretion rate. Therefore we further investigate the dependence of SFR and the star forming host fraction as a function of specific accretion rate, in order to remove any stellar mass dependence.
The specific accretion rate, $\lambda \propto L_{bol}/\mathcal{M}_*$, traces the rate of accretion scaled relative to the host stellar mass.  We calculate the specific accretion rate as

\begin{equation}  \label{eq:lambda}
\lambda=\frac{L_{bol}}{1.3\times 10^{38} \rm{erg \; s}^{-1}\times 0.002\frac{\mathcal{M}_*}{\mathcal{M}_\odot }}.                
\end{equation}
Thus, $\lambda$ is a rough tracer of the Eddington ratio, under the assumption that  $\mathcal{M}_*\approx \mathcal{M}_{bulge}$ and ${\mathcal{M}_{BH} \approx  0.002  \mathcal{M}_{bulge}}$ \citep{marconi2003relation}.

Figure \ref{fig:sfr_lambda} shows the SFR versus specific accretion rate of our sample, split into star forming and quiescent host galaxies. Both populations host AGN with a wide range of specific accretion rate, though the average specific accretion rate in star forming host galaxies is higher than in quiescent host galaxies in all three redshift panels of Figure \ref{fig:sfr_lambda}. 
However, the average stellar mass of the star forming galaxies is lower than that of the quiescent galaxies, and we can not detect lower specific accretion rate sources in lower stellar mass host galaxies, because of the X-ray flux limit. To minimize this bias, we first estimate an approximate X-ray luminosity limit in each redshift bin by taking the X-ray luminosity that 90\% of the sources exceed ($L_{90}$). 
This luminosity limit gives a rough indication of the luminosity below which we are no longer sampling the X-ray AGN population (for a given redshift) and is primarily determined by our deepest field (i.e. the CDFS)\footnote{We note this does \emph{not} correspond to the X-ray completeness corrections calculated with the full X-ray sensitivity curves that are described and applied in Section \ref{sec:pledd} to accurately recover the fraction of galaxies that host an AGN of a given specific accretion rate}.
This luminosity limit corresponds to a different limit in specific accretion rate, depending on the mass of the host. 
To convert to a specific accretion rate limit, we find the stellar mass above which 90\% of our X-ray sources lie ($\mathcal{M}_{90}$) in each redshift bin.
We convert $L_{90}$ and $\mathcal{M}_{90}$ to a limit in specific accretion rate, $\lambda_{90}$, using Equation \eqref{eq:lambda}.
We note that $\mathcal{M}_{90}$ is higher than our nominal stellar-mass-completeness limits (see Section \ref{sec:masscomp}) due to the fact that our X-ray sources are predominantly found in moderately massive hosts; thus our specific accretion rate limits are \emph{lower} than if our sample all had hosts with masses at the nominal stellar-mass-completeness limit.
Above our specific accretion rate limit, we should have a sample that is representative of the X-ray AGN sample.
 However, we will miss any \emph{lower mass} galaxies with the same accretion rate.

\begin{figure*}
\center
\includegraphics[width=0.7\textwidth]{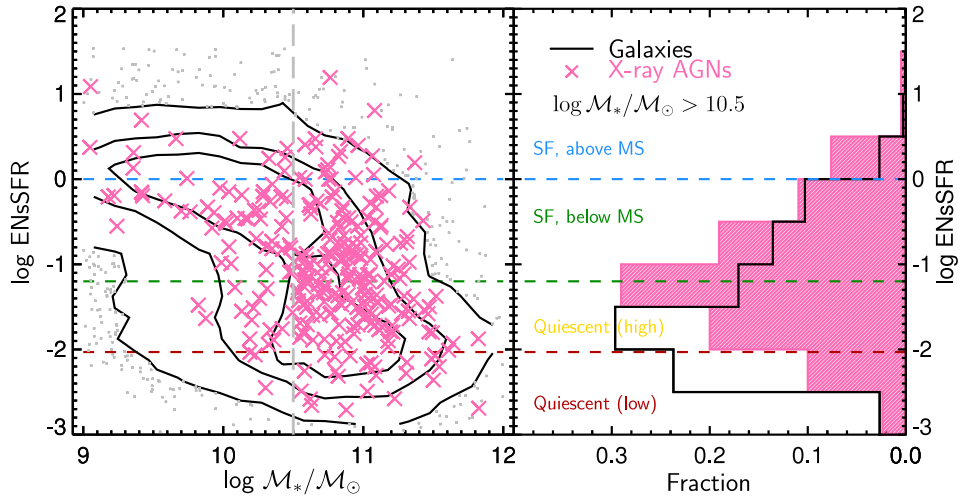}
\caption{
\textit{Left:} Epoch-normalized specific star formation rate ($ENsSFR$) versus stellar mass ($\mathcal{M}_*$) for the galaxy sample (black contours) and X-ray AGN (pink crosses) samples considered in Section \ref{sec:pledd}. 
The horizontal dashed lines indicate the dividing lines between our four $ENsSFR$ bins, as labeled. 
\textit{Right:} Normalized distributions of $ENsSFR$ for the galaxy and X-ray AGN samples, restricted to a stellar mass limit of $\log \mathcal{M}_*/\mathcal{M}_\odot > 10.5$ (indicated by the vertical dashed grey line in the left panel), where the majority of our X-ray AGN sample are detected and we are complete across our entire redshift range.
At these stellar masses the galaxy population is dominated by quiescent galaxies, but the X-ray AGN population is biased towards higher $ENsSFR$ sources.}
\label{fig:mstel_vs_enssfr_scatt_hist}
\end{figure*}

The shaded region in Figure \ref{fig:sfr_lambda} illustrates the range of specific accretion rates below this limit, where we will not have a representative sample. Above this limit we calculate the correlation coefficients of individual AGN, both the star forming and quiescent host populations separately, and find that the  correlation coefficients are negligible for quiescent host galaxies in all three redshift ranges. Within the star forming host population, in the lowest redshift range there is a weak negative correlation with a coefficient of -0.37 ($p$=0.03), such that an increase in the specific accretion rate corresponds to a decrease in the SFR.
This trend disappears in the two higher redshift panels.


\begin{table}
\caption{ The fraction of X-ray selected AGN with star forming host galaxies, and the fraction of stellar mass-matched galaxies that are star forming, above the specific accretion rate limit.}
\begin{center}
  \begin{tabular}{| c | c | c |}
    \hline
    Redshift& AGN with SF hosts & SF mass-matched galaxies \\
    \hline
    $0.2<z<0.5$ & 64$\pm$18\% & 54\% \\ \hline
    $0.5<z<0.8$ & 72$\pm$14\% & 45\% \\ \hline 
    $0.8<z<1.2$ & 55$\pm$21\% &  48\% \\
    \hline

  \end{tabular}
\end{center}
\label{tab:fraction_lambda}
\end{table}

Figure \ref{fig:fraction_lambda} shows the fraction of AGN with a star forming host galaxy as a function of AGN specific accretion rate. Similar to Figure \ref{fig:fraction}, this fraction is compared for X-ray AGN host galaxies, all galaxies above the stellar mass limit of the PRIMUS survey, and for galaxy samples with the same stellar mass distribution as the AGN host galaxies.  The comparison is done only above the specific accretion rate limit at a given redshift where we will have a representative sample.
As before, the error bars on the fractions are from the binomial distribution and
 are equivalent to 1$\sigma$ uncertainties. Table \ref{tab:fraction_lambda} lists the fraction of AGN host galaxies that are star forming (over all values of $L_\mathrm{X}$), as well as the fraction of stellar mass-matched galaxies that are star forming, in each redshift range, above our specific accretion rate limit.

We find that the fraction of AGN with star forming host galaxies at $0.2<z<0.5$ and $0.8<z<1.2$ above our specific accretion rate limits are not significantly different ($<1\sigma$) to the fraction of star forming galaxies in the mass-matched galaxy sample. 
In the middle redshift range ($0.5<z<0.8$), there is a 1.9$\sigma$ difference. 
Figure \ref{fig:fraction_lambda} shows that the fraction of AGN in star forming host galaxies may increase with specific accretion rate in the lower redshift range.
In the higher two redshift ranges this fraction appears to be roughly constant.  
These trends are generally consistent with what was found in Figure \ref{fig:fraction}, as a function of $L_\mathrm{X}$, but our limited sample size means we do not have a high significance result and thus these trends should be treated with caution.

To summarize, when we re-examine trends with SFR and the specific accretion rate (rather than $L_\mathrm{X}$) within our X-ray AGN sample, we do not find any highly significant ($>3\sigma$) correlations that were previously missed. 
Furthermore, the $3\sigma$ correlation between SFR and $L_\mathrm{X}$ for quiescent host galaxies that was seen in the first panel of Figure \ref{fig:sfr_lx} is no longer found when we renormalize $L_\mathrm{X}$ by the stellar mass (Figure \ref{fig:sfr_lambda}).
We note that applying the specific accretion rate limits has reduced our sample size and thus could be why we no longer find a significant correlation.
However, overall we conclude that there is no evidence for a correlation between SFR and specific accretion rate for either star forming or quiescent host galaxies within the X-ray AGN population.

\begin{figure*}
\center
\includegraphics[width=0.8\textwidth, trim=30 0 0 0]{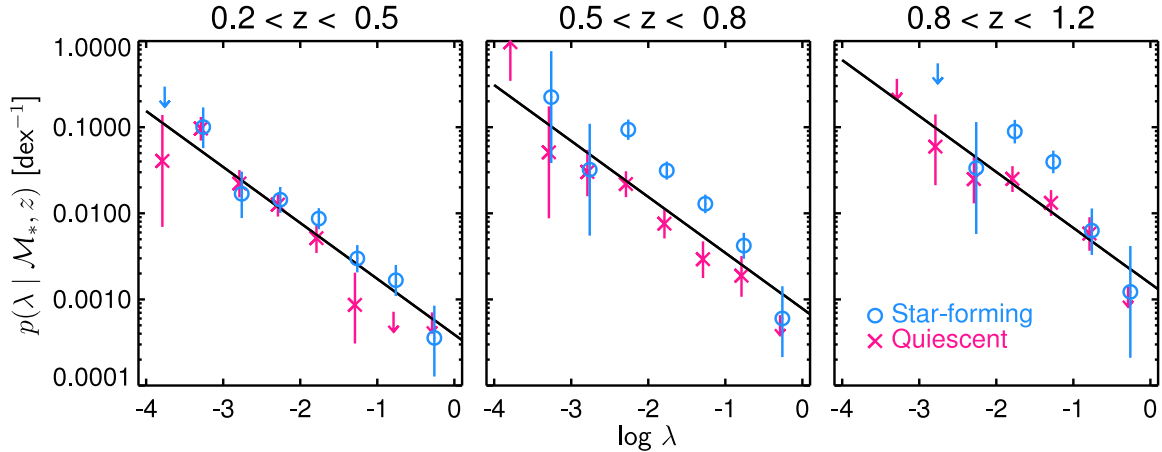}
\caption{
The probability density for a galaxy of given stellar mass, $\mathcal{M}_*$, and redshift, $z$, to host an AGN of specific accretion rate, $\lambda$, here dividing the galaxy sample into star forming and quiescent populations according to Equation 
\eqref{eq:enssfr}.
The thick black line corresponds to the best-fit power-law relationship measured by A12 for the overall galaxy population, evaluated at the center of the given redshift bin. Blue (red) points are estimated using the $N_\mathrm{obs}/N_\mathrm{mdl}$ method considering the star forming (quiescent) galaxy populations only, but with reference to this overall model that allows us to account for the underlying  redshift evolution, stellar-mass-dependent selection effects, and the X-ray completeness.
We see that at all redshifts the probability of a galaxy hosting an AGN depends strongly on specific accretion for both star forming and quiescent galaxy populations (rising towards low values of $\lambda$) but is a factor $\sim2-3$ higher in star forming galaxies than in quiescent galaxies at any given $\lambda$.
}
\label{fig:pledd_3zbins}
\end{figure*}

\subsection{The probability of a galaxy hosting an AGN as a function of star formation rate}
\label{sec:pledd}

In the above analysis we consider a sample of AGNs selected based on their X-ray emission, investigate the correlation between the SFR of the host galaxy and the AGN luminosity, and measure the fraction of these X-ray AGNs with hosts that are star forming versus quiescent.
In this section we take an alternative approach \citep[based on the approach of][hereafter A12]{aird2012primus}: we select samples of galaxies with a specified range of properties (in our case, a particular range of SFR) and determine the probability of finding an AGN in such galaxies. 
Our galaxy sample consists of all PRIMUS galaxies with stellar masses above the (redshift-dependent) completeness limits defined in Section \ref{sec:masscomp}. 
For this analysis we do not include galaxies or X-ray AGNs from the CDFS field; this field was a PRIMUS calibration field and the targeted galaxy sample is not defined in the same, consistent manner as the PRIMUS science fields. 
Thus, we are unable to accurately determine the probability of a galaxy hosting an AGN for the CDFS field.

We divide the galaxy sample according to the epoch-normalized specific SFR (ENsSFR), or the ratio of the sSFR to that of the main-sequence of star formation at the redshift of the galaxy, that we defined in Section \ref{sec:sf_and_q} (see Equation \eqref{eq:enssfr}).
By working with the $ENsSFR$ we can remove both the dependence of SFR on stellar mass and the overall evolution of the star forming main sequence over our redshift range. 
Figure \ref{fig:mstel_vs_enssfr_scatt_hist} (left panel) shows the distribution of $ENsSFR$ (as a function of stellar mass) for the galaxy and X-ray AGN samples we consider here, along with dividing lines between our populations. 
The horizontal green dashed line divides the quiescent and star forming galaxy populations, corresponding to the same $ENsSFR$ cuts used in Section \ref{sec:sf_and_q} ($\log $ENsSFR$ = -1.2$).
We also define two further divisions, corresponding to the peak of the star forming main sequence (at $\log ENsSFR=0$, by definition) and the peak of the quiescent galaxy population (at $\log ENsSFR=-2.01$).
These lines allow us to further sub-divide our galaxy sample into four populations: quiescent galaxies with low SFRs, quiescent galaxies with higher SFR, star forming galaxies below the star forming main sequence, and star forming galaxies above the star forming main sequence.

The right panel of Figure \ref{fig:mstel_vs_enssfr_scatt_hist} shows the distribution of $ENsSFR$ for the galaxy and X-ray AGN samples, above a mass limit of $\log \mathcal{M}_*/\mathcal{M}_\odot>10.5$ where the bulk of our X-ray detections lie and our galaxy sample is complete over the majority of the $0.2<z<1.2$ redshift range.
We note that at these masses there is no clear bimodality in the galaxy population.
Nonetheless, the X-ray AGNs show a distribution that is skewed towards higher $ENsSFR$ than that of the galaxies.
This appears consistent with the findings of Section \ref{sec:sf_and_q}, where we found weak evidence for a higher fraction of X-ray AGNs to be found in star forming galaxies (i.e. at higher ENsSFR) compared to galaxies of equivalent stellar mass.

To accurately determine the probability of a galaxy hosting an AGN we must correct for a number of sources of incompleteness in our X-ray selected AGN population.
These effects are ultimately due to the (varying) flux limit of the X-ray observations. 
The X-ray flux limit means that lower luminosity sources will not be identified at higher redshifts over the entire survey area, thus we must upweight any sources we \emph{do} detect. 
A12 also showed that probability of a galaxy hosting AGN is given by a power-law distribution of specific accretion rate ($\lambda$), where lower $\lambda$ sources are more common than high $\lambda$ sources in galaxies of all stellar masses \citep[see also][]{bongiorno2012accreting}.
However, as $\lambda$ scales with the host stellar mass, AGNs with the same $\lambda$ in a lower stellar mass host would have a lower observed X-ray luminosity, and thus may fall below our flux limits and be under-represented in our X-ray selected sample.

To account for these effects, we take the global relation from A12 for the probability density of a galaxy of given $\mathcal{M}_*$, and redshift, $z$, hosting an AGN of specific accretion rate, $\lambda$, per unit logarithmic $\lambda$, given by
\begin{equation}
p(\lambda \;|\; \mathcal{M}_*,z) \; d  \log L_\mathrm{X} = 
A \lambda^{\gamma_E} \left(\frac{1+z}{1+z_0}\right)^{\gamma_z} \; d \log \lambda
\end{equation}
where $z_0=0.6$ and the other parameters are given in table 3 of A12.
We use this relation, combined with the X-ray selection function, as a model to predict the observed number of X-ray AGN within one of our sub-samples of the galaxy population.
Thus, the predicted number of X-ray AGN within one of our sub-samples of the galaxy population is
\begin{equation}
N_\mathrm{mdl} = \sum_{k=1}^{N_\mathrm{samp}}
\int p(\lambda \;|\; \mathcal{M}_k,z_k) \;
p_\mathrm{det}\left(f_\mathrm{X}(\lambda,\mathcal{M}_k,z_k) \right)
\; d\log \lambda
\end{equation}
where the summation is performed over all $N_\mathrm{samp}$ galaxies in our sub-sample, $\mathcal{M}_k$ and $z_k$ are the stellar masses and redshifts of each galaxy, and $p_\mathrm{det}\left(f_\mathrm{X}(\lambda,\mathcal{M}_k,z_k) \right)$ is the probability of detecting an X-ray source with flux $f_\mathrm{X}$.

We calculate the X-ray flux,  $f_\mathrm{X}$, by converting $\lambda$ into a bolometric luminosity (given the stellar mass, $\mathcal{M}_k$, of the galaxy under consideration), which we then convert into an X-ray luminosity \citep[using the bolometric corrections of] []{hopkins2007observational}, and finally convert to an X-ray flux given the redshift of the galaxy, $z_k$. 
The dependence of $p_\mathrm{det}$ on flux is determined by the X-ray sensitivity curves, calculated in Section 4.1 of A12.
We use the ratio of this predicted number of X-ray AGNs to the actual observed number, $N_\mathrm{obs}$, in a given galaxy sub-sample to rescale the prediction for $p(\lambda \;|\; \mathcal{M}_*,z)$ from the A12 model at the centre of bin of given redshift and $\lambda$ \citep[based on the method of][]{miyaji2001soft}.
Errors on the binned estimates are based on the Poisson error given the observed number of X-ray AGNs \citep[taken from ][]{gehrels1986confidence}.
This method allows us to account for underlying variations of the model, the X-ray completeness, and the stellar mass and redshift distribution of the galaxy sample within a given bin.


\begin{figure}
\center
\includegraphics[width=0.9\columnwidth]{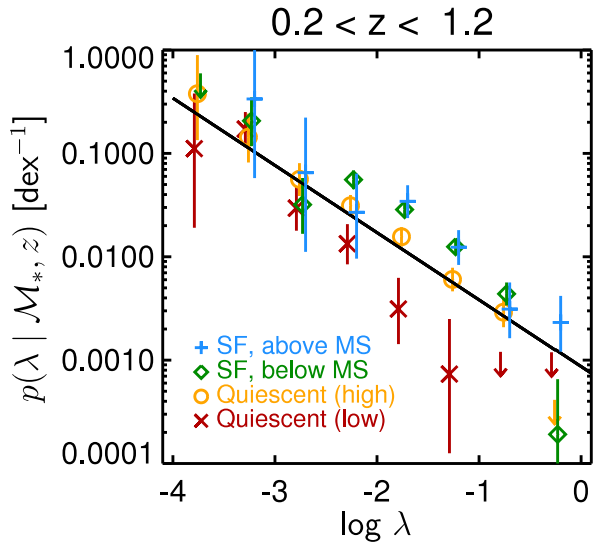}
\caption{
The probability density for a galaxy to host an AGN of specific accretion rate $\lambda$, dividing the galaxy sample into four populations according to their epoch-normalized specific star formation  rates (ENsSFRs). We consider a sample spanning the entire $0.2<z<1.2$ redshift range but account for the overall evolution in $p(\lambda \;|\; \mathcal{M}_*,z)$ according to the best-fit model of A12. The solid black line shows this best fit model evaluated at the center of the redshift bin($z=0.7$) and points are calculated using the $N_\mathrm{obs}/N_\mathrm{mdl}$ relative to this model, for each $ENsSFR$ bin. The shape of $p(\lambda \;|\; \mathcal{M}_*,z)$ remains roughly the same but the normalization increases moving from the quiescent populations to galaxies below or above the star forming main sequence (see also Figure \ref{fig:pledd_vs_enssfr}).
}
\label{fig:pledd_4enssfr_bins}
\end{figure}

\begin{figure}
\center
\includegraphics[width=0.85\columnwidth]{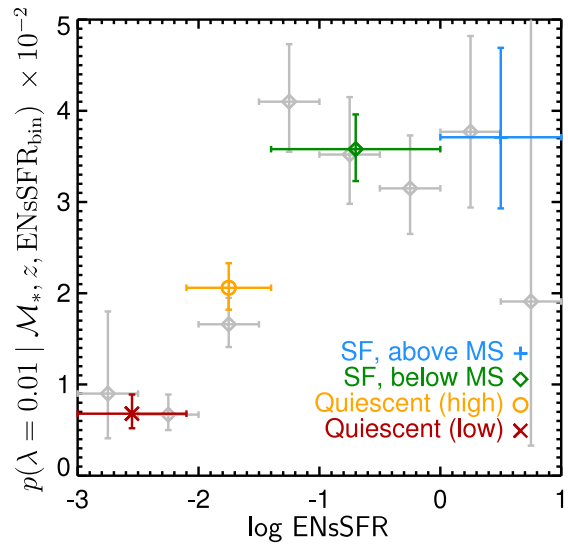}
\caption{
The probability density of a galaxy hosting an AGN, $p(\lambda \;|\; \mathcal{M}_*,z)$, evaluated at $\lambda=0.01$ and $z=0.7$, as a function of epoch-normalized specific star formation  rate ($ENsSFR$). Colored points correspond to bins defined relative to the star forming main sequence (as used in Figure \ref{fig:pledd_4enssfr_bins}), whereas the light grey points are for evenly spaced, 0.5 dex wide bins of ENsSFR.
The estimates use data from the entire $0.2<z<1.2$ redshift range and $-3<\log \lambda<-1$, but the plotted values are estimated at $\lambda=0.01$ and $z=0.7$ using the $N_\mathrm{obs}/N_\mathrm{mdl}$ method (relative to the A12 model).
We see that the probability of a galaxy hosting an AGN rises with $ENsSFR$ over the quiescent galaxy population but appears to flatten off for galaxies around the star forming main sequence.
}
\label{fig:pledd_vs_enssfr}
\end{figure}


In Figure \ref{fig:pledd_3zbins} we present binned estimates of $p(\lambda \;|\; \mathcal{M}_*,z)$ using the method described above for three redshift bins, dividing the galaxy sample into the quiescent (red crosses) and star forming (blue circles) populations based on our $ENsSFR$ cut. 
The black line indicates the underlying power-law model of A12.
We see a clear difference between the estimates for the star forming and quiescent populations at all redshifts, whereby the probability of finding an AGN (of fixed $\lambda$) is higher in a star forming galaxy than for a quiescent galaxy.
This difference is more significant at higher redshifts ($z>0.5$), where the star forming population lies above the global relation of A12. 
Both populations, however, appear to be consistent with the overall power-law shape of the distribution measured by A12; one type of galaxy population is not primarily associated with the most rapidly accreting sources. 
The overall increase in the probability of finding an AGN as redshift increases is also seen for both populations.

In Figure \ref{fig:pledd_4enssfr_bins} we combine our data over the entire $0.2<z<1.2$ redshift range but further subdivide our galaxy population into the four bins of $ENsSFR$ shown in Figure \ref{fig:mstel_vs_enssfr_scatt_hist}.
The underlying redshift evolution across this wide bin is corrected for by our $N_\mathrm{obs}/N_\mathrm{mdl}$ using the A12 model.
We again see that $p(\lambda \;|\; \mathcal{M}_*,z)$ has a fairly consistent power-law shape across the populations but the overall normalization increases moving across the quiescent population (from low to high SFR).

However, the normalization appears roughly constant across the star forming population.
To investigate these trends in more detail, we use our $N_\mathrm{obs}/N_\mathrm{mdl}$ method to estimate $p(\lambda \;|\; \mathcal{M}_*,z)$ for a wide $-3<\log \lambda<-1$ bin for each of the four populations (scaling relative to the A12 model evaluated at $\lambda=0.01$). 
We show these estimates as a function of $ENsSFR$ as the colored points in Figure \ref{fig:pledd_vs_enssfr}.
The light grey points also show estimates for 8 fixed width bins of $\log $ ENsSFR.
We confirm an increase (at $4 \sigma$ significance) in the normalization of $p(\lambda \;|\; \mathcal{M}_*,z)$ with $ENsSFR$ across the quiescent population, 
which increases further (by a further factor 1.7$\pm0.3$) to higher $ENsSFR$  (into the star forming galaxy population), where it appears to reach a plateau.

Overall, our results show that the probability of a galaxy hosting an AGN is higher for galaxies on the star forming main sequence. 
The probability of hosting an AGN drops for galaxies below the main sequence, where the star formation rates are lower. 
Nonetheless, AGNs are found in all types of galaxies and appear to have a similar overall distribution of accretion rates. Our results indicate that about 1--2\% of quiescent galaxies at $z\sim0.6$ host an AGN with an accretion rate of at least 1\% of Eddington. This rises in sources with higher SFRs and $\sim$ 3.5\% of star forming galaxies host an AGN with an accretion rate of at least 1\% of Eddington.

\section{Discussion}
\label{sec:discussion}

\subsection {Is there a correlation between SFR and AGN luminosity?}

In this paper we investigate the relationship between SFR and AGN luminosity, using 
$L_\mathrm{X}$, and generally find that these two quantities are not strongly correlated within moderate-luminosity X-ray selected AGN samples. There is a large scatter in SFR at any given $L_\mathrm{X}$ and more powerful AGN are 
not necessarily in more highly star forming galaxies.  We find evidence for a weak 
correlation 
between SFR and $L_\mathrm{X}$ in our full sample of both star forming and quiescent 
host galaxies at $z<0.8$. However, under further investigation we conclude this is likely 
due to 1) a weak correlation between SFR and $L_\mathrm{X}$ in low redshift 
quiescent host galaxies; we do not find a similar trend in star forming host galaxies at 
these redshifts, and/or 2) the fact that AGN with higher $L_\mathrm{X}$ are more 
likely to have a star forming host galaxy \citep[e.g.][]{kauffmann2003host,heckman2006host}, 
though our error bars are too large to measure this with significance.
We note that the SFR-$L_\mathrm{X}$ correlation observed in the full sample at $0.5<z<0.8$ vanishes when we investigate the correlation for star forming and quiescent host galaxies separately. This
further indicates that correlations seen for the full sample may be due to the changing mix of
star forming and quiescent hosts as a function of $L_\mathrm{X}$, rather than a correlation between 
SFR itself and $L_\mathrm{X}$. We confirm that none of the observed trends between SFR and $L_\mathrm{X}$ are driven by an underlying correlation between $L_\mathrm{X}$ and stellar mass within our AGN sample. While we find that the mean AGN host stellar mass is $ \log \frac {\mathcal{M}_*}{\mathcal{M}_\odot } \sim$10.5, there is a wide distribution in host galaxy stellar mass throughout our $L_\mathrm{X}$ range, consistent with \cite[e.g.][]{aird2012primus, bongiorno2012accreting.}
We also find a large scatter in specific accretion rate for both star forming and quiescent host galaxies, which could explain the lack of a direct correlation between SFR and $L_\mathrm{X}$.  In fact, for the low redshift quiescent host galaxies where we do find a significant correlation between SFR and $L_\mathrm{X}$, we do not find a correlation between SFR and specific accretion rate.

Overall, within the sample of AGN with star forming host galaxies, we do not find significant correlations between SFR and AGN luminosity. This is consistent with an emerging picture where the instantaneous BH accretion rate \citep[e.g.][]{chen2013correlation,hickox2014black} is decoupled from the current rate of star formation, at least in galaxies hosting moderate luminosity AGN \citep[e.g.][]{mullaney2012goods,rosario2012mean,harrison2012no,rovilos2012goods}. 
A possible scenario for BH fueling in low to moderate luminosity AGN is that it is 
driven by stochastic infall of cold gas from circumnuclear regions \citep[e.g.][]{kauffmann2009feast}. In fact star formation in most ``main sequence'' star forming galaxies appears to be driven by internal processes, e.g. disk instabilities and turbulence \citep[e.g.][]{elbaz2007reversal,daddi2010different}. Therefore, rather than there being a direct connection between star formation and AGN activity, these two processes likely share a common gas supply. 
However, there are studies that find a correlation between SFR and $L_\mathrm{X}$ in star forming galaxies that host powerful AGN, both at at $z<1$  \citep[e.g.][]{rosario2012mean} and $z>1$ \citep[e.g.][]{rovilos2012goods,netzer2013star}. There is also some evidence in LIRGS that indicates highly star forming galaxies are more likely to host AGN  \citep[e.g.][]{iwasawa}. As major merger events likely play an important role in fueling the most luminous AGN and starburst galaxies, then the incidence of major mergers could lead to a more direct correlation between SFR and AGN luminosity at high $L_\mathrm{X}$.

Our sample at $0.2 < z < 1.2$ also includes quiescent galaxies, for which we are able to 
estimate SFRs. This contrasts with studies that rely on far-IR data from \textit{Herschel}, as quiescent galaxies are not detected at these redshifts.
Interestingly, we find a significant correlation between SFR and $L_\mathrm{X}$ in quiescent host galaxies at $z<0.5$. This correlation could be interpreted as evidence of a direct coupling between SFR and  $L_\mathrm{X}$ in quiescent galaxies, for example due to a common cold gas supply. It could illustrate a different triggering mechanism (e.g. minor mergers or other secular processes) than that in star forming galaxies, which channels cold gas into the central regions.

We note that as our results rely on UV-optical SED fits, we could underestimate the level of dust (and therefore SFR) in a fraction of our galaxies. This could potentially introduce a bias that might hide an underlying correlation. Additionally, our SED templates do not include an AGN component; however, we have removed sources with broad lines in their optical spectra, and for our sample the light is clearly dominated by the host galaxy.  If there was any residual blue light from an AGN, this would decrease our estimated SFRs.  Unaccounted for, this could potentially introduce an observed correlation between SFR and $L_\mathrm{X}$ that does not exist; this could be happening in our low redshift quiescent host galaxies. Finally, our sample is limited to some extent by statistics and a larger and deeper sample would be helpful.

\subsection {Average X-ray luminosity versus SFR}

\begin{figure}[ht!]
\includegraphics[width=0.5\textwidth]{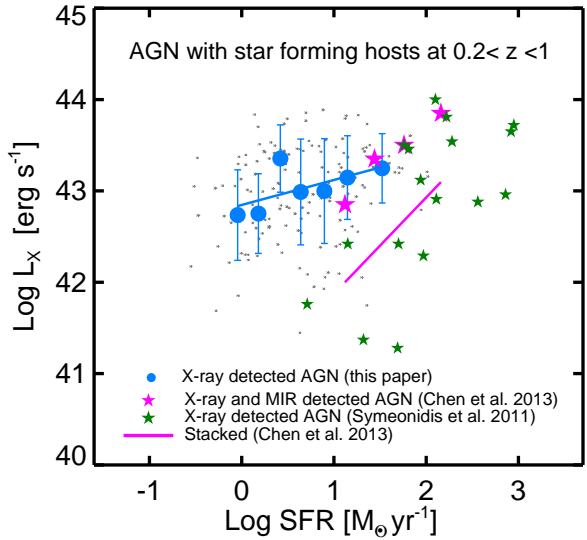}
\caption{{X-ray luminosity versus SFR in star forming host galaxies at redshift $0.2<z<1.0$. Gray points are PRIMUS individual X-ray detected AGN, while blue circles show the average $L_\mathrm{X}$  in bins of  SFR.  The error bars on each median point reflect the standard deviation. The blue solid line shows the best fitted line to the median points. 
We find a weak positive correlation with the coefficient of 0.24 ($p$= 0.00) between the average  $L_\mathrm{X}$ and SFR in our AGN sample. 
The median redshift in each bin of SFR indicates our correlation is not due to evolution of SFR with redshift. The pink stars are the detected AGN in \cite{chen2013correlation}, while the pink solid line indicates their average sample which also includes X-ray stacks of all galaxies. The green stars are X-ray detected AGN from  \cite{symeonidis2011herschel}, showing LIRG and ULIRG galaxies at $z\sim 1$. 
}}
\label{fig:average_lx_sfr}
\end{figure}

While we do not find a direct correlation between SFR and instantaneous X-ray luminosity in 
star forming host galaxies, these processes could still be connected through a common triggering and fueling mechanism. 
Due to a stochastic fueling process, accretion activity onto the SMBH is time variable \citep[e.g.][]{ulrich1997variability,peterson2001variability} and AGN luminosities may drop more than $10^2$ in less than $10^5$ years  \citep{keel2012history}. Additionally, recent studies found [OIII] emission from ionized clouds in the outskirt of galaxies with little or no evidence of on-going AGN activity \citep[e.g.][]{schawinski2010galaxy,keel2012galaxy,keel2012history}, confirming that AGN illuminating these clouds were much brighter in the past.
In contrast, star formation remains stable for a longer period of time; even in starburst galaxies this process can last $\sim$100 Myrs \citep[e.g.][]{wong2009timescale,hickox2012laboca}. Due to the rapid variability of AGN, using an instantaneous AGN luminosity to compare with SFR may hide an underlying connection between AGN and star formation. Instead, using an average AGN luminosity may be more appropriate to explore relationships between AGN and their host galaxies  \citep[e.g.][]{chen2013correlation,hickox2014black}. Recently, \cite{chen2013correlation} used the average AGN luminosity and found that the SMBH accretion rate is directly linked to the SFR in star forming galaxies at $z\sim 0.5$.

Here we use our sample of 167 X-ray detected AGN hosted by star forming galaxies at $0.2<z<1$ to consider the relation between the average $L_\mathrm{X}$ and SFR. We also compare our results with \cite{chen2013correlation} (hearafter C13). Since we use SED fits with UV and optical bands to estimate SFRs and C13 use \textit{Herschel}/SPIRE 250 $\mu m$ measurements, we use Equation \eqref{eq:lir} to convert their estimated IR luminosity to SFR for comparison. Figure \ref{fig:average_lx_sfr} shows $L_\mathrm{X}$ for our sample plotted as a function of SFR. Gray points indicate individual X-ray detected AGN, while blue circles show the average $L_\mathrm{X}$ of the gray points in bins of  SFR. Error bars on each median point reflect the standard deviation in that bin. The blue solid line shows the best fitted line to our median points, using a non-linear least square fit. We emphasize that this line is for X-ray detected AGN only.  

Pink stars show detected AGN in the C13 sample, which includes 34 X-ray detected AGN (using \textit{Chandra} ACIS-I 5 ks observations with $L_{X  0.5-7kev}$) and 87 MIR detected AGN  (using 4.5$\mu m$ observation from \textit{Spitzer}) in star forming galaxies in the Bo\"{o}tes field at $0.25 < z < 0.8$. From AGN detected in both X-rays and the MIR, C13 
used the median $\frac{L_\mathrm{X}}{L_{IR}}$ to estimate the $L_\mathrm{X}$ for MIR detected AGN. The solid pink line shows the average $L_\mathrm{X}$ found for both detected and undetected AGN, where C13 
stacked the X-ray emission around active and non-active galaxies, removing the contribution 
from XBs in the X-ray stack.
The green stars show X-ray detected AGN in  \cite{symeonidis2011herschel}, 
which includes 17 LIRG and ULIRG galaxies at $z\sim 1$,  from the 2 Ms observations in the CDFN.

In our detected AGN sample, there is a slight but significant trend here with a correlation coefficient of 0.24 ($p$= 0.00) between the average $L_\mathrm{X}$ and SFR. This is somewhat shallower than the trend in the C13 detected AGN sample, but where the samples overlap there is good agreement.  C13 are not able to detect AGN with $L_\mathrm{X} \lesssim 10^{43} $ erg s$^{-1}$ due to their shallower X-ray data, therefore their points do not probe to as low SFR as our sample. Since the total X-ray area of PRIMUS is smaller than the area probed by C13, and BLAGN are not included in this study, our sample lacks the highly star forming galaxies above $\log(\frac{SFR}{\mathcal{M}_\odot yr^{-1}}) \sim$ 1.5. 

We note that the trends shown in C13 for both the detected and average samples are affected by redshift. Once corrected for the evolution of the X-ray luminosity function and SFR with redshift, C13 find a shallower trend that is highly consistent with our result. To investigate whether evolution may have an effect on the trend observed in our sample, we find the median redshift in each bin of SFR.  We do not have a significant redshift-dependence in our SFR bins, such that our correlation is not impacted by evolution. 
X-ray selected AGN from \cite{symeonidis2011herschel} confirm the correlation between $L_\mathrm{X}$ and SFR found in C13, although this sample only includes AGN in higher SFR sources (LIRGs and ULIRGs). Furthermore deeper X-ray data may provide greater dynamic range in $L_\mathrm{X}$ revealing a steeper trend. Our sample includes lower SFR sources but it does not  probe as deep in X-ray luminosity and therefore is difficult to compare directly with the \cite{symeonidis2011herschel} sample.

Overall, our sample of X-ray detected AGN shows a large scatter in $L_\mathrm{X}$ 
at a given SFR, though we find a weak but significant correlation between them.  
This correlation indicates that the rate of black hole growth is related to the SFR 
in star forming galaxies, when effectively averaging black hole growth over long 
timescales, consistent with stochastic fueling of the AGN from the same ultimate 
fuel supply as that for star formation.
While here we averaged only our X-ray detected sources, ideally we would want to take 
the average over the entire galaxy sample. To do this properly we would need to  stack 
our galaxy sample, which is beyond the scope of this paper. We also note that the X-ray flux limit of our survey impacts the correlation that we find and that with deeper X-ray data we would be able to investigate this correlation more precisely.

\subsection{Where do AGN live?}

In our study, we find a large scatter between SFR and $L_\mathrm{X}$, with little evidence of a direct correlation, when considering X-ray selected AGN with either star forming or quiescent host galaxies. However, our results from Section \ref{sec:pledd} indicate that, when considering the entire galaxy population, one is more likely to find an AGN in a star forming galaxy.
Within either the star forming or quiescent galaxy populations, we find AGNs with a wide range of specific accretion rates, described by a roughly power-law distribution. 
However, for a given $\lambda$, the probability of a star forming galaxy hosting an AGN is higher than for a quiescent galaxy.  
Enhanced AGN activity in star forming galaxies has also been seen in several recent studies \citep[e.g.][]{silverman2009environments, mullaney2012goods,aird2012primus,rovilos2012goods,rosario2012mean}. The differences in the distributions can either be interpreted as an increased probability of AGN activity being triggered in galaxies with large reservoirs of cold gas (that also fuel star formation), or that AGNs in such galaxies are accreting, \emph{on average}, at higher rates \citep[see also][]{georgakakis2014investigating}.

These findings are consistent with the picture discussed above, where the level of AGN accretion in a given galaxy can vary substantially over short time periods (relative to the star formation timescales), and could explain that lack of a strong, direct correlation between SFR and $L_\mathrm{X}$:
while the overall probability of hosting an AGN is higher for higher SFRs, the instantaneous accretion rate that we observe from a single galaxy can vary over many orders of magnitude, washing out any direct correlation between the SFR and $L_\mathrm{X}$. 

While this is appealing, it is important to note where our results do not fit in with this simple picture.
Firstly, we do find AGNs in quiescent galaxies that may have very low levels of star formation and the distribution has a similar power-law shape (albeit shifted to lower $\lambda$), indicating that the underlying physical processes that regulate AGNs may be similar in quiescent galaxies to those taking place in star forming galaxies (although ultimately the large scale fueling processes may be different).
Secondly, we do not find a rise in the probability of hosting an AGN with SFR \emph{within} the star forming galaxy population itself, indicating that increased star formation does not go hand-in-hand with increased (average) BH growth.
Conversely, for quiescent galaxies with reduced SFRs --placing them below the main sequence of star formation-- 
we find that the probability of hosting an AGN is decreased by a factor $\sim2-3$. 
Furthermore, as SFRs decrease \emph{within} the quiescent population we find that the probability of hosting an AGN also decreases, indicating that as star formation is shut down there is also a reduction of AGN activity in quiescent galaxies. 
Nevertheless, as emphasized above, AGNs are still widespread within quiescent galaxies, with a wide range of specific accretion rates.

We note that the number of bins we used to classify our host galaxies is limited by our sample size, and with the current binning we do not have a sufficiently large sample to directly measure the shape of $p(\lambda \;|\; \mathcal{M}_*,z)$ within each bin. 
Larger samples would allow us to accurately track changes in the distribution of specific accretion rate as a function of the host galaxy properties and would shed light on any change in the underlying physical processes.

\section{Summary}
\label{sec:summary}

In this paper we study the relationship between AGN X-ray luminosity and their host galaxies' SFR and stellar mass. We use a sample of 309 X-ray selected AGN with spectroscopic redshifts from the PRIMUS survey at $0.2<z<1.2$. We exclude BLAGN to minimize the contribution of AGN light when estimating host galaxy properties, and we include AGN with $10^{41}<L_\mathrm{X}<10^{44} $ erg s$^{-1}$.
Our main conclusions are as follows:
\begin{itemize}

\item Star formation rate and AGN luminosity are not strongly correlated within our X-ray AGN sample at $0.2<z<1.2$.  There is a wide range of SFRs at a given $L_\mathrm{X}$, and a higher $L_\mathrm{X}$ does not necessarily imply a higher SFR.  We do not find any significant correlation between SFR and $L_\mathrm{X}$ in star forming host galaxies, though we do find a weak but significant correlation between the mean $L_\mathrm{X}$ of detected AGN and SFR. This correlation implies an underlying connection that may exist due to a common gas supply but the variability of AGN accretion on relatively short timescales makes it hard to observe.

\item AGN with a wide range of $L_\mathrm{X}$ reside in both star forming and quiescent galaxies with a wide range of stellar masses, although are generally found in moderately massive ($\gtrsim10^{10} \mathcal{M}_\odot$) galaxies.  However, we do not find any correlation between stellar mass and  $L_\mathrm{X}$ within our X-ray AGN sample for either the star forming or quiescent host populations.

\item We find a wide range of specific accretion rates, $\lambda$ ($L_\mathrm{X}$ normalized by host stellar mass), across the star forming and quiescent host populations, which could explain the lack of a stronger correlation between SFR and $L_\mathrm{X}$. 

\item The fraction of AGN residing in star forming host galaxies increases with increasing AGN X-ray luminosity, indicating that more powerful AGN are mainly hosted in star forming galaxies at $z<1$.

\item Finally, we consider the fraction of AGN within the entire galaxy population. The probability that a galaxy of a given stellar mass, $\mathcal{M}_*$, and redshift, $z$ hosts an AGN as a function of specific accretion rate, $p(\lambda \;|\; \mathcal{M}_*,z)$, is roughly a power law for both star forming and quiescent host galaxies.  The probability of hosting an AGN at a given specific accretion rate is higher for star forming galaxies than quiescent galaxies.
Furthermore, this probability increases with SFR within the quiescent galaxy population, though within the star forming population there is no change across the ``main sequence'' of star formation. 

\end{itemize}

Within star forming galaxies, known to contain abundant cold gas, we find no direct correlation between SFR and instantaneous AGN activity, although the overall probability of hosting an AGN is higher than in quiescent galaxies.
Conversely, in quiescent host galaxies, where the overall probability of finding an AGN is somewhat lower, we do find evidence for a correlation between SFR and AGN instantaneous luminosity which may suggest different triggering and fueling processes (e.g. minor mergers, secular processes) drive both star formation and AGN activity in such galaxies.
However, the distribution of accretion rates in both star forming and quiescent galaxies has a similar approximately power-law form, indicating that AGN accretion is ultimately a stochastic process and that the same physical processes may regulate AGN activity once gas is funneled to the central few parsecs.

\acknowledgements

We thank the referee for their positive comments and constructive advice which helped improving this paper. We also thank Ramin Skibba and Aleks Diamond-Stanic for helpful discussions and comments on the paper. We acknowledge Rebecca Bernstein, Adam Bolton, Douglas Finkbeiner, David W. Hogg, Timothy McKay, Sam Roweis, and Wiphu Rujopakarn for their contributions to the PRIMUS project. 

We thank the CFHTLS, COSMOS, DLS, and SWIRE teams for their public data releases and/or access to early releases for the PRIMUS survey. This paper includes data gathered with the 6.5m Magellan Telescopes located at Las Campanas Observatory, Chile. We thank the support staff at LCO for their help during our observations, and we acknowledge the use of community access through NOAO observing time. Some of the data used for this project are from the CFHTLS public data release, which includes observations obtained with MegaPrime/MegaCam, a joint project of CFHT and CEA/DAPNIA, at the Canada- France-Hawaii Telescope (CFHT) which is operated by the National Research Council (NRC) of Canada, the Institut National des Science de l'Univers of the Centre National de la Recherche Scientifique (CNRS) of France, and the University of Hawaii. This work is based in part on data products produced at TERAPIX and the Canadian Astronomy Data Centre as part of the Canada- France-Hawaii Telescope Legacy Survey, a collaborative project of NRC and CNRS. We also thank those who have built and operate the Chandra and XMM-Newton X-ray observatories. Funding for PRIMUS has been provided by NSF grants AST-0607701, 0908246, 0908442, and 0908354, and NASA grant 08-ADP08-0019. 
MA and ALC acknowledge support from NSF CAREER award AST-1055081.  
J.A. acknowledges support from a COFUND Junior Research Fellowship from the 
Institute of Advanced Study, Durham University.

\bibliographystyle{apj}
\bibliography{references}

\end{document}